\documentclass[11pt,a4paper]{article}
\usepackage[margin=2.5cm]{geometry}
\usepackage{amsmath,amssymb,amsfonts}
\usepackage{amsthm}
\newtheorem{proposition}{Proposition}
\usepackage{graphicx}
\usepackage{booktabs}
\usepackage{xcolor}
\usepackage[T1]{fontenc}
\usepackage{lmodern}

\usepackage[colorlinks=true,
            linkcolor=blue!60!black,
            citecolor=blue!60!black,
            urlcolor=blue!60!black,
            pdftitle={Charged Deformation of a Hairy Black Brane: Perturbative Solution and Thermodynamics},
            pdfsubject={Holographic charged black brane perturbation theory}
            ]{hyperref}

\newcommand{\eps}{\epsilon}
\newcommand{\order}[1]{\mathcal{O}\!\left(#1\right)}
\newcommand{\AdS}{\mathrm{AdS}}

\newcommand{\sE}{\mathcal{E}}
\newcommand{\src}{\varkappa}

\begin{document}

\begin{center}
{\LARGE\bfseries Charged Deformation of a Hairy Black Brane:\\[4pt]
Perturbative Solution and Thermodynamics}

\vspace{2cm}
{\large Sangheon Yun}\footnote{\texttt{sangheon.yun@gmail.com}}

\vspace{1.5cm}
\small IndigoWave, Center for Quantum Spacetime, Sogang University,\\
\small 35 Baekbeom-ro, Mapo-gu, Seoul 04107, Republic of Korea

\vspace{2cm}
\end{center}

\begin{abstract}
We study electrically charged black branes in four-dimensional Einstein-Maxwell-scalar theory with Maxwell coupling $Z(\phi)=e^{3\phi/\sqrt7}$, whose potential has an AdS$_4$ fixed point only in the infrared. For nonzero hair parameter $\alpha$ the scalar grows logarithmically at large radius and the geometry is not asymptotically AdS, so the holographic dictionary is a working assumption. With the scale-invariant multi-trace boundary condition of the neutral branes, giving conformal thermodynamics, we work to second order in the charge, obtaining the Maxwell field and susceptibility in closed form, plus the thermodynamics, condensate, incoherent conductivity and cumulant ratio $T^2\chi_4/\chi_2=6c(\alpha)<0$ for all $\alpha$ considered. Independent perturbative computations and first-law closed forms agree within $4\times10^{-6}$; nonlinear solutions reproduce second-order results to $1.2\times10^{-5}$ and $c$ to $6\times10^{-4}$. They are locally thermodynamically stable where constructed, and the pressure is analytic in $(\mu_Q/T)^2$ near zero density. No regular extremal horizon exists at finite scalar value, and $\rho/s<\sqrt{21}/(2\pi)$ exactly. The extremal limit is an AdS$_2\times\mathbb R^2$ throat where the scalar grows as $\ln\ln(1/u)$ in the AdS$_2$ radial coordinate $u$; the bound is saturated there, and extremal values are approached as powers of $1/\ln(\mu_Q/T)$. With a Dirichlet condition on the scalar source, the same solutions describe a non-conformal theory with scale $\Lambda$, whose charged branes exist only below $T_{\max}(\mu_Q)$, which falls from $0.194\Lambda$ at zero density to zero at $\mu_Q=0.506\Lambda$. This boundary is a diffusive instability where the charge susceptibility diverges at nonzero density, and the bulk viscosity violates Buchel's bound at all densities. Quasinormal modes with the normalizable large-radius condition reproduce its hydrodynamics.
\end{abstract}

\vspace{2cm}

\vspace{1.8cm}
\tableofcontents
\vspace{1.5cm}

\section{Introduction}
\label{sec:intro}

\subsection{Logarithmically running scalars and holography}

Deforming a holographic conformal field theory (CFT)~\cite{ref:Maldacena,ref:GKP,ref:Witten} by an operator dual to a bulk scalar triggers a renormalization-group (RG) flow, described in the bulk by a radially dependent scalar profile~\cite{ref:FGPW,ref:GPPZ,ref:PS-holored}; such flows have also been studied at finite temperature~\cite{ref:freedman-gpu}.
A classically marginal operator is a borderline case, in which backreaction induces a logarithmic running of the coupling~\cite{ref:Park}.

In this paper we study an Einstein--scalar theory in four dimensions with a two-exponential potential $V(\phi)$ in which a logarithmic running is realized in a different way.
The potential has an $\AdS_4$ fixed point at $\phi=0$, which governs the infrared.
In the ultraviolet, however, the scalar grows logarithmically with the radius, with a coefficient fixed by the potential, and for nonzero hair the geometry is not asymptotically $\AdS$: the curvature vanishes and the asymptotic region lies at infinite light-crossing time (Section~\ref{sec:setup}).
What survives of the marginal structure is a scaling symmetry under which the scalar source shifts additively, the analogue of the logarithmic running of a marginal coupling.
Standard holographic tools derived for asymptotically $\AdS$ spacetimes are therefore adopted here as explicit working assumptions, supported by internal consistency (Section~\ref{sec:setup}).

The neutral Einstein--scalar system arises as a consistent truncation of eleven-dimensional supergravity~\cite{Gauntlett:2009zw} or of type IIA supergravity on $\AdS_4\times\mathbb{CP}^3$~\cite{Bak:2010yd}, the background dual to ABJM theory~\cite{ref:ABJM}; the further truncation to the single neutral scalar with the potential~\eqref{eq:potential} considered here is given in Eqs.~(28)--(29) of Ref.~\cite{Yun:2010}.
The exact neutral black brane was found in Ref.~\cite{ref:CMS}, and its thermodynamics, including the restoration of conformal invariance in the infrared, was analyzed in Ref.~\cite{Cadoni:2012uf}.
Reference~\cite{ref:original} identified it as a thermal solution of the ABJM-dual truncation and analyzed its holographic dictionary, a scale-invariant multi-trace boundary condition under which the equilibrium thermodynamics is conformal, the resulting two-parameter thermodynamics and the stability of the neutral branes. Since the source mode is not normalizable, that reference regards the scale-invariant condition as an ensemble established only in equilibrium, and the theory with fixed source as the dual of the solutions.
The solution is parametrized by a mass parameter $m$ and a dimensionless hair parameter $\alpha\ge0$; for $\alpha=0$ it reduces to the planar $\AdS_4$-Schwarzschild hole, and the non-$\AdS$ ultraviolet region appears only for $\alpha>0$.

\subsection{Charged extensions and motivation}

Coupling a Maxwell field to this background is natural: a conserved global $U(1)$ current of the boundary theory is dual to a bulk gauge field, whose near-boundary expansion determines the chemical potential and the charge density.
Charged black branes underlie holographic models of finite-density matter~\cite{ref:HHH,ref:Hartnoll-review}, and charged dilatonic branes and their transport properties have been studied extensively~\cite{ref:GubserRocha,ref:GKPT,ref:CGKKM}.
Most of these studies involve scalars with power-law falloff; much less is known about finite-density states when the scalar source runs logarithmically in the sense described above.
Finite-density states of the ABJM-dual truncation have been studied with a $U(1)$ chemical potential in Ref.~\cite{ref:BakKimYun}, and the Maxwell coupling used here was proposed as the natural charged extension of the present background in Ref.~\cite{ref:original}.
No closed-form electrically charged extension is known: the Maxwell stress backreacts on both the metric and the marginal scalar.
The Maxwell field at first order in the charge, however, can be integrated in closed form, because $Z(\phi_0)=\Omega_0^3$ on the neutral solution (Section~\ref{sec:firstorder}).
We therefore first work perturbatively in the charge and then construct the solutions at arbitrary density numerically.
(Certain magnetically charged toroidal models, by contrast, admit an integrable reduction to deformed Toda equations~\cite{ref:YunToda}; see Refs.~\cite{Yun:2026uhb,Yun:2026dsh} for related perturbative expansions.)

In Sections~\ref{sec:setup}--\ref{sec:finite} we use the scale-invariant boundary condition of Ref.~\cite{ref:original}, which fixes the hair parameter $\alpha$ and makes the equilibrium thermodynamics conformal; the transport coefficients and hydrodynamic poles obtained there are therefore those of hydrodynamics built on this equilibrium ensemble, and Section~\ref{sec:dir:qnm} shows which of them are realized as quasinormal modes. In Section~\ref{sec:dir} we apply the charged solutions to the theory with fixed source of the same reference, which is non-conformal.
Its boundary potential is unbounded below, in close analogy with conformally invariant triple-trace deformations of the M2-brane theory~\cite{ref:HH-designer,ref:HH-cosmo,ref:CHT}, so that the thermal states are at best metastable (Section~\ref{sec:thermo:stability}).

\subsection{Main results and structure}

The neutral branes, their thermodynamics, the boundary condition and the stability of the neutral branes were studied in Refs.~\cite{ref:CMS,Cadoni:2012uf,ref:original}. What is new here is the electric charge, treated both perturbatively and nonlinearly.
The central part of this work is the second-order perturbation theory in the charge, $\eps^2\equiv Q^2/l^2\ll1$; the nonlinear solutions at finite density serve as an independent check and extend the results beyond the perturbative regime. Our main results are:
(i)~a formulation of the $\order{\eps^2}$ problem in variables regular at the horizon, a characterization of its three-parameter family of regular solutions, and a labeling at fixed entropy and fixed hair parameter;
(ii)~the extension of the boundary condition of Ref.~\cite{ref:original} to finite charge: imposing the same coupling $k$ directly through the Ward identity~\eqref{eq:full_k}, the second-order and nonlinear solutions reproduce the first-law predictions, so that no further freedom enters;
(iii)~the complete set of $\order{\eps^2}$ observables: the temperature, entropy, energy and pressure shifts, fixed in closed form by the first law in terms of the susceptibility, together with the shifts of the marginal source, the condensate and the horizon scalar and the charge correction to the incoherent conductivity (Table~\ref{tab:thermo});
(iv)~the third-order chemical potential, which fixes the quartic term of the pressure and gives the cumulant ratio $T^2\chi_4/\chi_2=6c(\alpha)$, negative for all values of $\alpha$ considered (Section~\ref{sec:second});
(v)~a verification of every second-order quantity by independent methods, together with first-law closed forms (Table~\ref{tab:second_checks});
(vi)~the nonlinear solutions at arbitrary density, which confirm the second-order results and reveal the extremal behavior: no regular extremal horizon exists at finite scalar value, and $\rho/s$ obeys the exact bound $\rho/s<\sqrt{21}/(2\pi)$; and
(vii)~the extremal infrared throat, which is locally $\AdS_2\times\mathbb R^2$ with a scalar that grows as $\ln\ln(1/u)$ in the $\AdS_2$ radial coordinate $u$. It saturates the bound and fixes the logarithmic approach of the low-temperature observables to their extremal values (Proposition~\ref{prop:throat}); and
(viii)~the charged thermodynamics of the theory with fixed source (Section~\ref{sec:dir}): the susceptibility and the second-order thermodynamics in closed form, a quartic coefficient that diverges at the maximal temperature, the region of existence below $T_{\max}(\mu_Q)$ and its stability, the transport coefficients, the zero-temperature equation of state, and quasinormal modes that confirm the hydrodynamic predictions and show that the instability at the boundary of the stable region is diffusive.
In addition, we show that the charged solutions are analytic in the charge density.
A common thread is that $Z(\phi)$ and the leading term of $V(\phi)$ carry opposite exponentials, which controls the physics at finite density, from the sign of the quartic charge response to the extremal throat, where it produces a flat direction for the scalar and the saturation of the bound on $\rho/s$.

The paper is organized as follows.
Section~\ref{sec:setup} reviews the neutral background and sets up the charged perturbation.
Sections~\ref{sec:firstorder}--\ref{sec:horizon} solve the Maxwell equation at first order, derive the second-order system, analyze horizon regularity and express the entropy and temperature in terms of horizon data.
Section~\ref{sec:thermo} summarizes the boundary condition and develops the charged thermodynamics, Section~\ref{sec:transport} derives the transport coefficients, Section~\ref{sec:second} collects the second-order data and verifies it, and Section~\ref{sec:finite} constructs the nonlinear solutions at finite density.
Section~\ref{sec:dir} treats the theory with fixed source.
Section~\ref{sec:discussion} summarizes the results and lists open problems; the appendices treat the near-horizon structure, the extremal throat, the asymptotic region for time-dependent perturbations and the numerical convergence.

\section{Setup and neutral background}
\label{sec:setup}

We consider the four-dimensional action
\begin{equation}
I = \frac{1}{16\pi G_4}\int d^4x\,\sqrt{-g}
    \left[\mathcal{R} - \tfrac{1}{2}(\partial\phi)^2 - V(\phi)
          - Z(\phi)\,F_{\mu\nu}F^{\mu\nu}\right],
\qquad Z(\phi) = e^{\tfrac{3}{\sqrt7}\phi},
\label{eq:action}
\end{equation}
with the two-exponential scalar potential
\begin{equation}
V(\phi) = -\frac{21}{2l^2}\,e^{-\tfrac{3}{\sqrt7}\phi}
          + \frac{9}{2l^2}\,e^{-\sqrt7\,\phi}.
\label{eq:potential}
\end{equation}
The neutral sector is the one studied in Refs.~\cite{ref:CMS,Cadoni:2012uf,ref:original}.
The Maxwell coupling $Z(\phi)$ is motivated by the same reduction of type IIA supergravity: in the four-dimensional action of Refs.~\cite{Bak:2010yd,Yun:2010} (Eq.~(13) of Ref.~\cite{Yun:2010}), the Ramond--Ramond one-form couples to $\phi$ through $e^{3\phi/\sqrt7}$ when the second scalar $\phi_-$ of that reference vanishes.
A truncation that retains this vector alone is not consistent there, because it has a St\"uckelberg mass term and its field strength sources $\phi_-$; we therefore regard~\eqref{eq:action} with this $Z(\phi)$ as a model in its own right.
Two of its properties play a central role below: $Z(\phi_0)=\Omega_0^3$ on the neutral solution, which makes the first-order Maxwell problem solvable in closed form (Section~\ref{sec:firstorder}), and $Z$ carries the exponential opposite to that of the leading term of $V$, which controls the extremal limit (Section~\ref{sec:finite:thermo}).

In a Schwarzschild-like radial gauge, the exact neutral ($Q=0$) black brane is
\begin{equation}
ds^2=-a_0(r)^2\,dt^2+b_0(r)^2\,dr^2+h_0(r)^2\,(dx^2+dy^2),
\end{equation}
with
\begin{align}
a_0(r)^2 &= f_0(r)\,\Omega_0(r),&
h_0(r)^2 &= r^2\,\Omega_0(r),&
b_0(r)^2 &= \frac{\Omega_0(r)^{3}}{f_0(r)},
\label{eq:neutral_metric}\\[4pt]
f_0(r)   &= \frac{r^2}{l^2} - \frac{2m}{r},&
\Omega_0(r)^2 &= \frac{\sinh\alpha}{ml^2}\,r^3 + e^{-\alpha},&
\phi_0(r) &= \frac{\sqrt7}{2}\ln\Omega_0(r)^2,
\label{eq:neutral_profile}
\end{align}
where $m > 0$ is the mass-density parameter (all densities are per unit area of the $(x,y)$ plane) and $\alpha \ge 0$ is the hair parameter.
We write $w(r)\equiv\Omega_0(r)^2$ below.
The horizon is located at $r_h = (2ml^2)^{1/3}$, where $f_0(r_h)=0$ and $w(r_h) = e^\alpha$, and the temperature and entropy density are~\cite{Cadoni:2012uf,ref:original}
\begin{equation}
T^{(0)} = \frac{3\,r_h\,e^{-\alpha/2}}{4\pi l^2},\qquad
S^{(0)} = \frac{r_h^2\,e^{\alpha/2}}{4G_4},\qquad
T^{(0)}S^{(0)}=\frac{3m}{8\pi G_4}.
\label{eq:neutral_thermo}
\end{equation}
In Ref.~\cite{ref:original} the neutral thermodynamics was organized in terms of the mass $M^{(0)}=m/(4\pi G_4)=\tfrac23 T^{(0)}S^{(0)}$ and an extended first law involving a thermodynamic central charge $C_t\propto e^{3\alpha/2}$~\cite{ref:Visser,ref:CKM}.
In Section~\ref{sec:thermo} we derive $M^{(0)}$ as the energy density of the dual theory with the scale-invariant boundary condition on the marginal scalar.

\paragraph{Asymptotic structure.}
At $\alpha=0$ the scalar vanishes and the brane is the planar $\AdS_4$-Schwarzschild solution.
For $\alpha>0$, which is the case of interest here, the only $\AdS_4$ region of the theory is the infrared fixed point $\phi=0$ of the potential, which is reached as $r\to0$ by the zero-temperature domain wall of Section~\ref{sec:thermo:stability} and is never reached by the black branes, and the large-$r$ region is not asymptotically $\AdS$.
The crossover to this region occurs at $r\sim\src^{-1/3}$, which recedes to infinity as $\alpha\to0$; this non-uniformity underlies the non-analytic small-$\alpha$ behavior found in Section~\ref{sec:firstorder}.
At $\alpha=0$ the ultraviolet is $\AdS_4$ and $\phi$ is dual to an irrelevant operator with $\Delta=6$ (Section~\ref{sec:thermo:stability}).
Because $Z'(0)\neq0$, the Maxwell field sources $\phi$ even at $\alpha=0$, and the charged problem then has a different asymptotic structure and different boundary conditions.
The $\alpha=0$ values quoted below, such as $j(0)=1$ and $D_Q=3/(4\pi T)$, are first-order results for a probe Maxwell field on $\AdS_4$-Schwarzschild; the second-order and nonlinear results of this paper apply to $\alpha>0$ only.
For $r\to\infty$ one has $w\simeq\src r^3$ with $\src\equiv\sinh\alpha/(ml^2)$ (see~\eqref{eq:varphi1}), and the metric becomes
\begin{equation}
ds^2\simeq\src^{1/2}\,r^{7/2}\Bigl(-\frac{dt^2}{l^2}+dx^2+dy^2+d\zeta^2\Bigr),\qquad
\zeta\equiv2\,l\,\src^{1/2}\,r^{1/2},
\label{eq:UVmetric}
\end{equation}
up to relative corrections of order $r^{-3}$.
Three properties follow.
First, the curvature vanishes instead of approaching the $\AdS_4$ value $-12/l^2$.
Using the trace of the Einstein equations, $\mathcal R=\tfrac12(\partial\phi)^2+2V$, one finds $\mathcal R\simeq-\tfrac{105}{8}\,l^{-2}\src^{-3/2}r^{-9/2}\to0$, while $V(\phi_0)\to0^-$ because $\phi_0\to+\infty$.
Second, the metric is conformal to flat space with a conformal factor that grows as $\zeta^7$, and the asymptotic region $\zeta\to\infty$ lies at infinite light-crossing time, $\int b_0/a_0\,dr\propto r^{1/2}$. It is therefore not a timelike conformal boundary, and there is no Fefferman--Graham expansion about an $\AdS$ metric.
Equivalently, the asymptotic metric has the hyperscaling-violating form with dynamical exponent $z=1$ and $\theta=9$ in the conventions of Refs.~\cite{ref:CGKKM,ref:DHKTW}, as noted in Ref.~\cite{ref:original}.
Hyperscaling violation for this class of scalar black branes was identified in Ref.~\cite{Cadoni:2012uf}, where the corresponding parameter of the scalar-dressed phase is reported to be negative, which indicates a different convention for $\theta$.
Third, the scalar grows logarithmically, $\phi_0\simeq\varphi_{(0)}+\varphi_{(1)}\ln(r/l)$, with a coefficient $\varphi_{(1)}=3\sqrt7/2$ that is fixed by the potential rather than by the source (Section~\ref{sec:thermo:source}).
Accordingly, in this paper ``marginal'' refers to the following structure: under the scaling symmetry of the neutral family the source shifts additively, $\varphi_{(0)}\to\varphi_{(0)}-\varphi_{(1)}\ln\lambda$, which is the analogue of the logarithmic running of a marginal coupling. It does not refer to the relation $m_\phi^2l^2=\Delta(\Delta-d)$ at an ultraviolet $\AdS$ fixed point, which does not exist here.
It is related to the characterization of Ref.~\cite{ref:original}, where the operator is called marginally relevant because the effective mass $V''(\phi_0)$ approaches $0^-$ at large $r$; holographic RG flows of classically marginal operators whose ultraviolet fixed point lies at $\phi\to\pm\infty$ have been analyzed in Ref.~\cite{ref:BourdierKiritsis}.
The asymptotic structure resembles that of non-conformal branes with a generalized conformal structure~\cite{ref:KST,ref:KanitscheiderSkenderis}, but differs from it in that the asymptotic region lies at infinite light-crossing time.

\paragraph{Working assumptions.}
Several standard holographic tools have been derived and tested for asymptotically (locally) $\AdS$ spacetimes, and because of the structure just described they are not guaranteed to apply here. As in Ref.~\cite{ref:original}, we adopt them as working assumptions:
\begin{itemize}
\item[(A1)] \emph{Holographic renormalization.} The renormalized on-shell action and stress tensor of the dual theory are given by the Hamilton--Jacobi counterterm of Refs.~\cite{ref:dBVV,ref:PS} with the exact superpotential of Section~\ref{sec:thermo:holoren}, and the Dirichlet source and response are $\varphi_{(0)}$, defined invariantly through~\eqref{eq:src_inv}, and $\langle\mathcal O_\phi\rangle$. The general results of Refs.~\cite{ref:dHSS,ref:BFS} rely on a Fefferman--Graham expansion about an asymptotically $\AdS$ metric and do not cover the present case.
\item[(A2)] \emph{Multi-trace interpretation.} The boundary term $\mathcal W(\varphi_{(0)})$ of Section~\ref{sec:thermo:conformal} can be interpreted, as in $\AdS$~\cite{ref:Witten-multitrace,ref:HMTZ,ref:Papadimitriou-MT}, as a multi-trace deformation of the dual theory, i.e.\ the Legendre-transform structure of the generating functional carries over. Because the asymptotic region is not a timelike boundary, we use this identification only for static configurations (Appendix~\ref{app:dynamics}).
\item[(A3)] \emph{Energy bound.} The criterion of Hertog and Horowitz~\cite{ref:HH-designer}, formulated and tested for asymptotically $\AdS$ spacetimes with scalars that admit two normalizable falloffs, also governs the lower bound of the energy here.
\end{itemize}
The evidence for (A1) and (A2) is internal consistency.
The divergences cancel, and the resulting energy, pressure, condensate and free energy satisfy the first law, the trace Ward identity and the Smarr-type identity, both for the neutral branes and at finite charge, to the accuracies reported in Sections~\ref{sec:thermo}, \ref{sec:second:checks} and~\ref{sec:finite:checks}.
These checks are nontrivial, but they are not a derivation.
The conclusions of Section~\ref{sec:thermo:stability} on the unboundedness of the energy rely in addition on (A3) and are conditional on it.
The bulk solutions themselves, the first-order Maxwell field and the horizon quantities (entropy, surface gravity and the horizon formulas for transport) do not depend on (A1)--(A3).

\paragraph{Charged extension.}
We switch on a bulk electric field $A = A_t(r)\,dt$ coupled through $Z(\phi)$ and work perturbatively in $\eps \equiv Q/l \ll 1$:
\begin{equation}
A_t(r) = \eps\,A_t^{(1)}(r) + \order{\eps^3},\qquad
\phi(r) = \phi_0(r) + \eps^2\,\delta\phi(r) + \order{\eps^4},
\label{eq:pert_fields}
\end{equation}
with the metric functions expanded analogously (Section~\ref{sec:secondorder}).
For any quantity $X$ we write $X = X^{(0)}+\eps^2X^{(2)}+\order{\eps^4}$.
The physical dimensionless expansion parameter is the ratio of chemical potential to temperature, $\mu_Q/T = \eps\,\mu_Q^{(1)}/T^{(0)}+\order{\eps^3}$, with $\mu_Q^{(1)}$ given in~\eqref{eq:muQ}.

\section{First order: the Maxwell field}
\label{sec:firstorder}

The Maxwell equation $\partial_\mu\!\big(\sqrt{-g}\,Z(\phi)\,F^{\mu\nu}\big)=0$ has the first integral $\sqrt{-g}\,Z(\phi)\,F^{tr} = Q$.
At leading order in $\eps$ the background values $\sqrt{-g}=r^2\Omega_0^3$ and $Z(\phi_0)=\Omega_0^3$ give $A_t'=Q/(r^2\Omega_0^2)$, that is,
\begin{equation}
A_t^{(1)\prime}(r) =
\frac{m l^3\,e^{\alpha}}{r^2\bigl(m l^2 + r^3 e^{\alpha}\sinh\alpha\bigr)}.
\label{eq:At1prime}
\end{equation}
Defining $\mathfrak{b} \equiv (e^{2\alpha}-1)^{-1/3}$ and $z \equiv r/r_h$, this integrates to
\begin{align}
A_t^{(1)}(r) &= \frac{l e^\alpha}{r_h}
  \Bigl[\bigl(1-\tfrac{1}{z}\bigr) -F(z,\mathfrak{b})+F(1,\mathfrak{b})\Bigr],
\label{eq:At1closed}\\
F(z,\mathfrak{b}) &= \frac{1}{6\mathfrak{b}}\left[-\ln\!\bigl(9\mathfrak{b}^2(\mathfrak{b}+z)^2\bigr)
                             +\ln\!\bigl(\mathfrak{b}^2-\mathfrak{b}z+z^2\bigr)
                             -2\sqrt3\,\arctan\!\!\left(\frac{\sqrt3\,(\mathfrak{b}-2z)}{3\mathfrak{b}}\right)
                      \right],
\label{eq:Fdef}
\end{align}
normalized so that $A_t^{(1)}(r_h)=0$.
The chemical potential is
\begin{equation}
\mu_Q \equiv \eps\,A_t^{(1)}(\infty)
      = \eps\,\frac{l\,e^\alpha}{r_h}\,j(\alpha)\equiv\eps\, \mu_Q^{(1)},\qquad
j(\alpha) \equiv \int_1^\infty\!\frac{dz}{z^2\bigl[1+(e^{2\alpha}-1)z^3\bigr]}
= 1+F(1,\mathfrak{b})-F_\infty(\mathfrak{b}),
\label{eq:muQ}
\end{equation}
with $F_\infty(\mathfrak{b}) = [-\ln(9\mathfrak{b}^2)+\sqrt3\,\pi]/(6\mathfrak{b})$.
We have checked the closed form against direct numerical quadrature of~\eqref{eq:At1prime} to eight significant figures.
At $\alpha=0$ one has $j=1$ and $\mu_Q^{(1)}=l/r_h$, the $\AdS_4$-Schwarzschild value. For small $\alpha$ the approach is non-analytic,
\begin{equation}
j(\alpha) = 1 - \frac{2\pi}{3\sqrt3}\,(2\alpha)^{1/3} + \alpha + \order{\alpha^{4/3}},
\label{eq:j_small}
\end{equation}
because the integral in~\eqref{eq:muQ} is dominated by $z\sim(e^{2\alpha}-1)^{-1/3}$; the analytic term $+\alpha$ comes from the lower limit $z=1$.

The charge density conjugate to $\mu_Q$ is
\begin{equation}
\rho \equiv \frac{Q}{4\pi G_4\, l^2}=\frac{\eps}{4\pi G_4\, l}.
\label{eq:Qth}
\end{equation}
The factor $1/(4\pi G_4)$ reflects the normalization of the Maxwell term in~\eqref{eq:action}.
Writing the Maxwell Lagrangian as $-\tfrac14\,[Z/(4\pi G_4)]\,F^2$ shows that $Z/(4\pi G_4)$ plays the role of the effective inverse squared gauge coupling.
Since $\mu_Q = 4\pi G_4\,l\,\mu_Q^{(1)}\rho$ is linear in $\rho$ at this order, the charge susceptibility is
\begin{equation}
\chi \equiv \frac{\partial\rho}{\partial\mu_Q}\bigg|_{T} = \frac{1}{4\pi G_4\, l\,\mu_Q^{(1)}},
\label{eq:chi}
\end{equation}
known in closed form. As we show in Section~\ref{sec:thermo}, $\chi$ determines the entire $\order{\eps^2}$ thermodynamics.

\section{Second order: field equations and gauge fixing}
\label{sec:secondorder}

At order $\eps^2$ the Maxwell stress tensor backreacts on the metric and the scalar.
We write
\begin{equation}
a(r)^2 = a_0(r)^2\bigl[1+\eps^2\,\delta A(r)\bigr],\quad
h(r)^2 = h_0(r)^2\bigl[1+\eps^2\,\delta H(r)\bigr],\quad
b(r)^2 = b_0(r)^2\bigl[1+\eps^2\,\delta B(r)\bigr],
\label{eq:pert_ansatz}
\end{equation}
together with $\phi = \phi_0 + \eps^2\delta\phi$.
Expanding the field equations to $\order{\eps^2}$ yields four equations (the $tt$, $rr$ and $xx$ components of the Einstein equations and the scalar equation) for the three dynamical functions $\delta A$, $\delta H$ and $\delta\phi$, while $\delta B$ enters only algebraically.
The $rr$ equation contains at most first derivatives of the perturbations and can be solved for $\delta B$.
The resulting second-order system for $(\delta A,\delta H,\delta\phi)$ has a degenerate coefficient matrix of rank two: as dictated by the contracted Bianchi identity, only two of the three second-order equations are independent once the constraint and the matter equations hold.
The redundancy reflects the residual diffeomorphism $r \to r + \eps^2\xi(r)$, which we fix by choosing
\begin{equation}
\delta H(r) \equiv 0,
\label{eq:gauge}
\end{equation}
so that $h(r) = h_0(r)$ holds exactly at order $\eps^2$.
The remaining system has total differential order four.
A convenient independent set consists of the $tt$ and $rr$ Einstein equations, which are first order in $\delta B$ and $\delta A$, respectively, and the second-order scalar equation; the $xx$ equation then follows, as we have verified along our numerical solutions.

\section{Horizon regularity and construction of the solution}
\label{sec:regularity}

\subsection{Regular variables}
\label{sec:regularity:vars}

The perturbations $\delta A$ and $\delta B$ are singular at the neutral horizon: a charge-induced shift of the horizon, $r_h\to r_h+\eps^2\delta r_h$, appears as simple poles
\begin{equation}
\delta A(r) = -\frac{\delta r_h}{r - r_h} + D + \order{r-r_h},\qquad
\delta B(r) = +\frac{\delta r_h}{r - r_h} + D_B + \order{r-r_h},
\label{eq:NHexpand}
\end{equation}
with opposite residues (Appendix~\ref{app:A}). We therefore work with the variables
\begin{equation}
U(r)\equiv f_0(r)\,\delta A(r),\qquad g(r)\equiv \delta A(r)+\delta B(r),\qquad \delta\phi(r),
\label{eq:regvars}
\end{equation}
all of which are regular at $r_h$, with $U(r_h)=-f_0'(r_h)\,\delta r_h$ and $g(r_h)=D+D_B$.
In these variables the system consists of first-order equations for $U$ and $g$ and a second-order equation for $\delta\phi$.
The equations for $U'$ and $g'$ have coefficients that are finite at $r_h$; in particular the $tt$--$rr$ combination of the Einstein equations, to which a purely electric field does not contribute, gives
\begin{equation}
g'(r) = \frac{h_0(r)\,\phi_0'(r)}{h_0'(r)}\,\delta\phi'(r)
\label{eq:gprime}
\end{equation}
(Appendix~\ref{app:A}).
The scalar equation takes the form $\delta\phi''=\mathcal N(r;U,g,\delta\phi,\delta\phi')/f_0(r)$, with $\mathcal N$ linear in its arguments and regular at $r_h$.
Regularity of $\delta\phi$ therefore requires $\mathcal N(r_h)=0$, which fixes $\delta\phi'(r_h)$ as a linear function of $(U(r_h),g(r_h),\delta\phi(r_h))$ and excludes the logarithmically divergent mode.
The remaining regular solutions are analytic at the horizon: starting the integration at $r_h(1+x_0)$ with a first-order Taylor expansion, our results are independent of $x_0$ for $10^{-7}\le x_0\le10^{-5}$ to all quoted digits (Table~\ref{tab:conv}), which indicates the absence of $(r-r_h)\ln(r-r_h)$ terms.

\subsection{The space of regular solutions}
\label{sec:regularity:space}

A regular solution is thus specified by the three horizon data $(U(r_h),g(r_h),\delta\phi(r_h))$.
The regular solutions form a three-parameter affine family: a particular solution sourced by the Maxwell stress, plus arbitrary multiples of three regular homogeneous solutions.
These are
(a)~a constant rescaling of time, $\delta A=g=\text{const}$, $\delta\phi=0$;
(b) and (c)~the derivatives of the neutral family with respect to $m$ and $\alpha$, each supplemented by the compensating diffeomorphism that preserves $\delta H=0$.
Modes (b) and (c) merely relabel the neutral background; they do not change the physics but shift the values of $m$ and $\alpha$ assigned to a given charged solution.

At large $r$ every solution of the linearized system approaches constants, $\delta A\to\delta A_\infty$, $g\to g_\infty$ and $\delta\phi\to\delta\phi_\infty$, with corrections of order $r^{-3}$; in particular $r\,\delta\phi'\to0$ for all solutions.
The linearized system has no $\ln r$ mode, so the log-running coefficient $\varphi_{(1)}$ of Section~\ref{sec:thermo:source} cannot be shifted at this order: it is fixed by the potential.

The three labels are fixed by physical conditions, summarized in Table~\ref{tab:bc}.
The time normalization is fixed at the boundary, $\delta A_\infty=0$ (Section~\ref{sec:thermo:source} explains why this is the correct normalization).
One relabeling mode is fixed by holding the entropy fixed, $\delta r_h=0$. The other is fixed by the scale-invariant boundary condition on the marginal scalar (Section~\ref{sec:thermo:conformal}), which holds $\alpha$ fixed and thereby determines the $\order{\eps^2}$ shift $\delta\varphi_{(0)}$ of the marginal source.
In practice we first construct the solution with unshifted source, $\delta\phi_\infty=0$, and then add the relabeling mode at fixed entropy that shifts the source by the amount~\eqref{eq:dphi0} required by the boundary condition.
In this labeling, every $\order{\eps^2}$ quantity is a correction at fixed entropy $S$ and fixed $\alpha$, which we denote by $|_{S,\alpha}$.
Any other labeling gives physically equivalent results once expressed in terms of physical variables. We have checked explicitly that the temperature correction at fixed $(S,\src)$, where $\src$ is the source parameter defined in~\eqref{eq:varphi1}, is independent of the labeling to seven significant figures (Appendix~\ref{app:C}).

\begin{table}[ht]
\centering
\caption{The four conditions that fix the solution of the fourth-order system.}
\label{tab:bc}
\renewcommand{\arraystretch}{1.2}
\begin{tabular}{lll}
\toprule
Condition & Physical meaning & Implementation \\
\midrule
$\mathcal N(r_h)=0$ & regular scalar at the horizon & fixes $\delta\phi'(r_h)$ \\
$U(r_h)=0$ & fixed entropy, $\delta r_h=0$ & horizon datum \\
boundary condition~\eqref{eq:mixedBC} & fixed $\alpha$ (scale invariance) & source shift~\eqref{eq:dphi0} \\
$\delta A_\infty=0$ & boundary time normalization & combination of basis solutions \\
\bottomrule
\end{tabular}
\end{table}

\subsection{Numerical procedure}
\label{sec:regularity:numerics}

We construct the particular solution with horizon data $U=g=\delta\phi=0$ and the homogeneous solution with $U=g=0$, $\delta\phi(r_h)=1$, each by integrating outward from $r_h(1+x_0)$ with $x_0=10^{-7}$ to $r_{\max}=400\,l$ ($\approx317\,r_h$), using an LSODA integrator with relative tolerance $10^{-12}$.
The time-rescaling mode is known in closed form.
The combination with $\delta\phi_\infty=0$ and $\delta A_\infty=0$ is formed algebraically, and the relabeling to fixed $\alpha$ is then performed analytically with the neutral thermodynamics.
The asymptotic constants are read off at $r_{\max}$, where the $\order{r^{-3}}$ corrections affect the results at the $10^{-8}$ level (Appendix~\ref{app:C}).

\section{Entropy and temperature from horizon data}
\label{sec:horizon}

In the gauge $\delta H \equiv 0$ the entropy density depends on the charge only through the horizon shift,
\begin{equation}
\frac{S^{(2)}}{S^{(0)}} =
\frac{2\,\delta r_h\,h_0'(r_h)}{h_0(r_h)} = 2\beta(\alpha)\,\frac{\delta r_h}{r_h},
\qquad
\beta(\alpha) \equiv \frac{r_h\,h_0'(r_h)}{h_0(r_h)}= 1 + \tfrac{3}{2}\,e^{-\alpha}\sinh\alpha ,
\label{eq:S2formula}
\end{equation}
where we used $h_0'/h_0 = 1/r + w'/(4w)$, $w(r_h)=e^\alpha$, $w'(r_h)=3(\sinh\alpha/ml^2)\,r_h^2$ and $r_h^3 = 2ml^2$. The factor $\beta$ increases monotonically from $1$ at $\alpha=0$ to $7/4$ as $\alpha\to\infty$.

The temperature is set by the surface gravity and therefore depends on both $a^2$ and $b^2$. In the boundary time normalization $\delta A_\infty=0$ we find (Appendix~\ref{app:A})
\begin{equation}
\frac{T^{(2)}}{T^{(0)}}
 = D -\,3\,e^{-\alpha}\sinh\alpha\,\frac{\delta r_h}{r_h}
 -\frac{g(r_h)}{2},
\qquad D = \frac{U'(r_h)}{f_0'(r_h)},
\label{eq:T2formula}
\end{equation}
where $D$ is the constant term in the near-horizon expansion~\eqref{eq:NHexpand} of $\delta A$.
Under a constant rescaling of time, $D$ and $g(r_h)$ shift by the same amount $c$, and $T^{(2)}/T^{(0)}$ by $c/2$, as it must; this is why the normalization has to be fixed at the boundary rather than at the horizon.
Both~\eqref{eq:S2formula} and~\eqref{eq:T2formula} hold in any labeling.
In the labeling of Table~\ref{tab:bc}, $\delta r_h=0$, so $S^{(2)}=0$ and $T^{(2)}/T^{(0)}=D-g(r_h)/2$ is directly the temperature correction at fixed $(S,\alpha)$.

\section{Thermodynamics}
\label{sec:thermo}

\subsection{Boundary condition and neutral thermodynamics}
\label{sec:thermo:source}\label{sec:thermo:holoren}\label{sec:thermo:conformal}\label{sec:thermo:stability}

The dictionary, the boundary condition and the stability analysis of the neutral branes were derived in Ref.~\cite{ref:original}. We summarize the results used below in the notation of this paper.

Near the boundary the neutral scalar behaves as
\begin{equation}
\begin{gathered}
\phi_0(r) = \varphi_{(0)} + \varphi_{(1)}\ln(r/l) + \varphi_{(3)}(l/r)^3 + \cdots,\\
\varphi_{(0)}=\frac{\sqrt{7}}{2}\ln\bigl(\src\, l^3\bigr),\qquad
\varphi_{(1)} = \frac{3\sqrt{7}}{2},\qquad
\src\equiv\frac{\sinh\alpha}{ml^2},
\end{gathered}
\label{eq:varphi1}
\end{equation}
where $\varphi_{(1)}$ is the coefficient of the logarithmic running, fixed by the potential, and $\varphi_{(0)}$ is the source.
Since $\varphi_{(0)}$ depends on the choice of radial coordinate, we define the source invariantly through the leading asymptotics: with the boundary coordinates $(x,y)$ held fixed,
\begin{equation}
\src^{4/7} = \lim_{r\to\infty} e^{2\phi/\sqrt7}\,h^{-12/7}.
\label{eq:src_inv}
\end{equation}
The boundary time is normalized by requiring the boundary metric to be conformal to $-dt^2+dx^2+dy^2$, i.e.\ $a^2/h^2\to1/l^2$; in the gauge $\delta H\equiv0$ this is the condition $\delta A_\infty=0$ of Table~\ref{tab:bc}, and the source is shifted at $\order{\eps^2}$ by $\delta\varphi_{(0)}=\tfrac74\,\delta\phi_\infty$.
Under the scaling $r\to\lambda r$, $m\to\lambda^3 m$ at fixed $\alpha$, which maps neutral branes into one another with $S\to\lambda^2S$ and $T\to\lambda T$, the source shifts as $\varphi_{(0)}\to\varphi_{(0)}-\varphi_{(1)}\ln\lambda$, so that a Dirichlet condition on $\varphi_{(0)}$ breaks scale invariance.

Under assumption (A1), holographic renormalization with the Hamilton--Jacobi counterterm~\cite{ref:dBVV,ref:PS} built from the exact superpotential
\begin{equation}
\mathcal U(\phi)=\frac{7}{2l}\,e^{-\frac{3}{2\sqrt7}\phi}-\frac{3}{2l}\,e^{-\frac{\sqrt7}{2}\phi},\qquad
V(\phi)=2\,\mathcal U'(\phi)^2-\tfrac32\,\mathcal U(\phi)^2,
\label{eq:superpotential}
\end{equation}
which generates the zero-temperature domain wall $f_0=r^2/l^2$, $\Omega_0^2=\src r^3+1$~\cite{ref:Bremer,Yun:2010}, gives with Dirichlet conditions the quasilocal (Brown--York~\cite{ref:BrownYork,ref:BK}) stress tensor
\begin{equation}
T^{i}{}_{j} = \frac{1}{8\pi G_4}\Bigl[\,K\,\delta^{i}{}_{j} - K^{i}{}_{j} - \mathcal U(\phi)\,\delta^{i}{}_{j}\,\Bigr],
\label{eq:BYmixed}
\end{equation}
with energy density $\sE_{\rm D}\equiv-\sqrt{-\gamma}\,T^{t}{}_{t}$ and pressure $p_{\rm D}\equiv\sqrt{-\gamma}\,T^{x}{}_{x}$,
\begin{equation}
\sE_{\rm D} = \frac{m}{16\pi G_4}\Bigl(4 + 3\tanh\tfrac{\alpha}{2}\Bigr),
\qquad
p_{\rm D} = \frac{m}{16\pi G_4}\Bigl(2 - 3\tanh\tfrac{\alpha}{2}\Bigr)=-\mathfrak f_{\rm D},
\label{eq:eps_p}
\end{equation}
and the Dirichlet first law and trace Ward identity
\begin{equation}
d\sE_{\rm D}=T\,dS+\langle\mathcal O_\phi\rangle\,d\varphi_{(0)},\qquad
\langle\mathcal O_\phi\rangle = -\frac{3\sqrt7\,m}{56\pi G_4}\,\tanh\frac{\alpha}{2},\qquad
T^{i}{}_{i}=-\sE_{\rm D}+2p_{\rm D}=\varphi_{(1)}\,\langle\mathcal O_\phi\rangle .
\label{eq:condensate}
\end{equation}

The scale-invariant boundary condition of Ref.~\cite{ref:original} is
\begin{equation}
\langle\mathcal O_\phi\rangle=-\mathcal W'\bigl(\varphi_{(0)}\bigr),\qquad
\mathcal W(\varphi_{(0)})=k\,e^{-2\varphi_{(0)}/\sqrt7},
\label{eq:mixedBC}
\end{equation}
i.e.\ a multi-trace deformation~\cite{ref:HMTZ,ref:Papadimitriou-MT} (assumption (A2)) of weight $d=3$, hence scale invariant, and marginal at leading order in $1/N$. On the neutral family it holds with
\begin{equation}
k=-\frac{3\,l}{16\pi G_4}\,(\cosh\alpha-1),\qquad
\mathcal W=-\frac{3m}{16\pi G_4}\tanh\frac{\alpha}{2},
\label{eq:k_alpha}
\end{equation}
so that fixing the coupling $k$ is equivalent to fixing $\alpha$.
The energy density and pressure are then $\sE=\sE_{\rm D}+\mathcal W=m/(4\pi G_4)=M^{(0)}$ and $p=p_{\rm D}-\mathcal W=\tfrac12M^{(0)}$, the stress tensor is traceless, and the first law holds at fixed coupling, $dM^{(0)}=T\,dS$.
The equilibrium thermodynamics of the neutral branes is therefore conformal, with $\sE=2p$, $c_s^2=1/2$ and $M^{(0)}=\tfrac23T^{(0)}S^{(0)}$ for all $\alpha$, while the free energy of the Dirichlet theory differs from $\mathfrak f$ by $-\mathcal W$.
Varying the coupling gives $d\sE=T\,dS+e^{-2\varphi_{(0)}/\sqrt7}\,dk$, which is the extended first law of Ref.~\cite{ref:original} with the thermodynamic central charge $C_t = \frac{32l^2}{\pi G_4}\,e^{3\alpha/2}$ a function of $k$.
For the energy bound, the boundary effective potential of the neutral theory is $\mathcal V=\mathcal W=k\sigma^3$ with $\sigma\equiv e^{-2\varphi_{(0)}/(3\sqrt7)}$, since the zero-temperature solitons have $\sE_{\rm D}=p_{\rm D}=\langle\mathcal O_\phi\rangle=0$.
The black branes have $k<0$ for every $\alpha>0$ by~\eqref{eq:k_alpha}, so $\mathcal V\to-\infty$ as $\sigma\to\infty$.
Under assumption (A3), the Hertog--Horowitz criterion~\cite{ref:HH-designer} then implies that the energy is unbounded below, as for the conformal triple-trace deformations of Refs.~\cite{ref:HH-designer,ref:HH-cosmo,ref:CHT}.
At $\alpha=0$ one has $k=0$ and the energy is bounded below, so the unboundedness is specific to the hairy branes.
This global statement is distinct from the local properties of the branes: they have positive specific heat (Section~\ref{sec:finite:thermo}) and never reach the infrared fixed point $\phi=0$, since $\phi_0$ increases monotonically from $\phi_h=\sqrt7\alpha/2$.

\paragraph{Dynamical stability.}
By the correlated stability conjecture of Gubser and Mitra~\cite{ref:GM1,ref:GM2}, positive specific heat is a necessary condition for the dynamical stability of a translationally invariant black brane, but not a sufficient one.
A systematic search for growing homogeneous modes of the neutral branes in the coupled scalar--gravity channel was reported in Ref.~\cite{ref:original} and finds none in the region of the complex frequency plane and for the values of $\alpha$ examined; Appendix~\ref{app:dynamics} recalls the asymptotic structure that enters this analysis.
This is a consistency check rather than a proof of stability. At nonzero wavevector, however, the sound channel of the branes with $\alpha>\alpha_*=\tfrac12\ln5$ has a growing mode; it involves only normalizable perturbations and does not depend on the boundary condition imposed on $\varphi_{(0)}$ (Section~\ref{sec:dir:qnm}); for $\alpha<\alpha_*$ we found only damped hydrodynamic modes. For $\alpha<\alpha_*$, together with the unbounded boundary potential, these results support the interpretation of the branes as metastable states: they cannot be the global ground state at any temperature, and decay requires large excursions of $\sigma$ (conditional on (A3)).
We have not computed the decay rate.
With Dirichlet conditions the same branes describe a non-conformal theory with a limiting temperature and a bulk viscosity below Buchel's bound, as found in Ref.~\cite{ref:original}; its charged branes are the subject of Section~\ref{sec:dir}.

\subsection{Charge corrections from the first law}
\label{sec:thermo:charge}

At fixed coupling $k$ (i.e.\ fixed $\alpha$) the first law reads $d\sE=T\,dS+\mu_Q\,d\rho$, with $\sE=\sE_{\rm D}+\mathcal W$.
At fixed $(S,\rho)$ the boundary condition~\eqref{eq:mixedBC} is the stationarity condition $\partial(\sE_{\rm D}+\mathcal W)/\partial\varphi_{(0)}=0$.
The $\order{\rho^2}$ shift of $\varphi_{(0)}$ that it induces therefore does not contribute to $\sE$ at this order, and
\begin{equation}
\sE(S,\rho;\alpha)=M^{(0)}(S;\alpha)+\frac{\rho^2}{2\chi(S;\alpha)}+\order{\rho^4},
\qquad
\sE^{(2)}\big|_{S,\alpha}=\frac{\mu_Q^{(1)}}{8\pi G_4\,l},
\label{eq:E_SQ}
\end{equation}
with $\chi$ the closed-form susceptibility~\eqref{eq:chi}.
At fixed $\alpha$, $\mu_Q^{(1)}\propto1/r_h\propto S^{-1/2}$, so $\sE$ is homogeneous of degree $3/2$ in $(S,\rho)$, and Euler's theorem gives the charged Euler relation of the conformal fluid,
\begin{equation}
3\sE=2TS+2\mu_Q\rho .
\label{eq:euler}
\end{equation}
The temperature $T=\partial\sE/\partial S|_{\rho,\alpha}$ receives the correction
\begin{equation}
\boxed{\;\frac{T^{(2)}}{T^{(0)}}\bigg|_{S,\alpha}=\frac{1}{8\pi G_4\,l\,T^{(0)}}\,\frac{\partial\mu_Q^{(1)}}{\partial S}\bigg|_{\alpha}
=-\frac{\mu_Q^{(1)}}{16\pi G_4\,l\,S^{(0)}T^{(0)}}=-\frac{\mu_Q^{(1)}}{6\,m\,l}\;}
\label{eq:T2_closed}
\end{equation}
which is negative for all $\alpha$ and decreases monotonically in magnitude.
At fixed temperature, using $S^{(0)}\propto T^2$ at fixed $\alpha$,
\begin{equation}
\frac{S^{(2)}}{S^{(0)}}\bigg|_{T,\alpha}=-2\,\frac{T^{(2)}}{T^{(0)}}\bigg|_{S,\alpha}=\frac{\mu_Q^{(1)}}{3\,m\,l}>0 ,
\label{eq:S2_fixedT}
\end{equation}
and the pressure as a function of $(T,\mu_Q)$ is $p=p^{(0)}(T;\alpha)+\tfrac12\chi\,\mu_Q^2+\order{\mu_Q^4}$, with $\chi\propto T$ as required by conformal invariance.

\paragraph{Source shift.}
Although it does not affect $\sE$ at this order, the boundary condition does shift the source.
Expanding~\eqref{eq:mixedBC} at fixed $S$,
\begin{equation}
\delta\varphi_{(0)}\big|_{S,\alpha}=-\,\frac{\partial_{\varphi_{(0)}}\!\bigl(\rho^2/2\chi\bigr)}{\partial^2_{\varphi_{(0)}}\sE_{\rm D}^{(0)}+\mathcal W''}\bigg|_{S},
\label{eq:dphi0}
\end{equation}
where the derivatives are taken along the neutral family at fixed $S$ and $\mathcal W''=\tfrac47\mathcal W$.
This is the shift used to label the bulk solution (Table~\ref{tab:bc}); its values are listed in Table~\ref{tab:thermo}.

\subsection{Comparison with the bulk solution}
\label{sec:thermo:check}

The closed form~\eqref{eq:T2_closed} uses only the first-order gauge field and the first law.
The bulk solution determines the same quantity independently from the horizon geometry, through the surface gravity~\eqref{eq:T2formula}. We compare the two in two steps.
First, before the boundary condition is imposed, the solution with unshifted source must obey the Maxwell relation of the Dirichlet first law~\eqref{eq:condensate}, $T^{(2)}|_{S,\src}=(8\pi G_4 l)^{-1}\,\partial\mu_Q^{(1)}/\partial S|_{\src}$.
It does so to a relative accuracy of about $10^{-8}$ for all values of $\alpha$ considered (e.g.\ $T^{(2)}/T^{(0)}|_{S,\src}=-0.04734166$ at $\alpha=0.6$); this checks the second-order solution independently of the boundary condition.
Second, after the relabeling~\eqref{eq:dphi0} to fixed $\alpha$, the surface-gravity temperature agrees with~\eqref{eq:T2_closed} to a relative accuracy of $3\times10^{-8}$ for all ten values of $\alpha$ (Figure~\ref{ChargedScalarfig}a).
The complete second-order data are collected in Table~\ref{tab:thermo} of Section~\ref{sec:second}, and further independent checks are described there.
The residual is set by the finite outer radius $r_{\max}$ (Appendix~\ref{app:C}).

\paragraph{Horizon scalar.}
At fixed $(S,\alpha)$ both the source shift and the horizon value $\delta\phi(r_h)$ are negative and decrease monotonically in magnitude with $\alpha$ (Table~\ref{tab:thermo} and Figure~\ref{ChargedScalarfig}b).
The value of $\delta\phi(r_h)$ is meaningful only once the labeling of Table~\ref{tab:bc} is specified; in a different labeling it is shifted by the relabeling modes and need not be monotonic.

\section{Boundary hydrodynamics and transport coefficients}
\label{sec:transport}

The transport coefficients below are obtained from horizon data in the limit $\omega\to0$, in which the radiating asymptotic region $r\gtrsim1/(\src\omega^2)$ of Appendix~\ref{app:dynamics} recedes to infinity.
In the ensemble with fixed $\alpha$ the equilibrium states of the charged black brane are those of a $(2+1)$-dimensional conformal fluid at finite temperature and, at order $\eps$, at finite charge density $\rho$. Since this ensemble is established only in equilibrium~\cite{ref:original}, the hydrodynamics below is built on its equation of state together with transport coefficients from horizon data; which of the resulting poles are quasinormal modes is stated in Section~\ref{sec:transport:modes}.
In this section we extract the transport coefficients accessible from horizon data, using the membrane-paradigm and horizon methods of Refs.~\cite{ref:KSS,ref:Iqbal-Liu,ref:DonosGauntlett,ref:DGH} together with relativistic hydrodynamics~\cite{ref:HLS}.
All thermodynamic derivatives are taken at fixed $\alpha$, and we denote the entropy density by $s\equiv S$. We keep $G_4$ explicit in formulas and quote numerical values in units $16\pi G_4 = l = m = 1$.

\subsection{Viscosities and speed of sound}
\label{sec:transport:shear}\label{sec:transport:bulk}

For a planar black hole with transverse metric $h_0(r)^2(dx^2+dy^2)$, the graviton zero mode in the shear channel decouples and yields the universal Kovtun--Son--Starinets result~\cite{ref:KSS,ref:Buchel}
\begin{equation}
\frac{\eta}{s} = \frac{1}{4\pi},\qquad
\eta = \frac{r_h^2\,e^{\alpha/2}}{16\pi G_4},
\label{eq:KSS}
\end{equation}
for all $\alpha$, and at order $\eps^2$ the shear viscosity and the entropy receive the same fractional correction.
Since the equilibrium thermodynamics at fixed $\alpha$ is conformal (Section~\ref{sec:thermo:conformal}), $\sE=2p$ holds identically, including at finite density by~\eqref{eq:euler}, and
\begin{equation}
c_s^2 = \frac{\partial p}{\partial\sE}\bigg|_{s/\rho} = \frac{1}{2},\qquad
\zeta = 0 ,
\label{eq:zeta_zero}
\end{equation}
for all $\alpha$ and to all orders in $\eps$.
This agrees with the Eling--Oz formula~\cite{ref:EO}, $\zeta/\eta=\bigl(s\,\partial_s\phi_h+\rho\,\partial_\rho\phi_h\bigr)^2$: at fixed $\alpha$, scale invariance implies that $\phi_h$ depends on $(s,\rho)$ only through $\rho/s$, which is annihilated by the dilatation operator $s\partial_s+\rho\partial_\rho$.

\subsection{Conductivities of a clean charged fluid}
\label{sec:transport:sigma}

Because the background is translation invariant, momentum is conserved, and at nonzero density the DC electrical, thermoelectric and thermal conductivities are all infinite: each contains a Drude weight.
At zero wavevector, relativistic hydrodynamics gives~\cite{ref:HLS}
\begin{equation}
\sigma(\omega) = \sigma_Q + \frac{\rho^2}{\sE+p}\,\frac{i}{\omega},\qquad
\vartheta(\omega) = -\frac{\mu_Q\,\sigma_Q}{T} + \frac{s\,\rho}{\sE+p}\,\frac{i}{\omega},\qquad
\bar\kappa(\omega) = \frac{\mu_Q^2\,\sigma_Q}{T} + \frac{s^2\,T}{\sE+p}\,\frac{i}{\omega},
\label{eq:hydro_cond}
\end{equation}
where $\vartheta$ is the thermoelectric conductivity and each pole $i/\omega$ is accompanied by a delta function $\pi\delta(\omega)$ in the real part.
These expressions, like the hydrodynamic poles of Section~\ref{sec:transport:modes}, are predictions of relativistic hydrodynamics given the thermodynamics and the horizon transport coefficients derived here.
A bulk computation at finite frequency requires a definition of the source in an asymptotic region at infinite light-crossing time (Appendix~\ref{app:dynamics}). For the theory with fixed source we adopt one in Section~\ref{sec:dir:qnm}; there the form~\eqref{eq:hydro_cond} is reproduced, and the radiating region $r\gtrsim1/(\src\omega^2)$ couples to the horizon only at order $(\omega/\Lambda)^9$, which bounds the additional dissipation it can contribute. We have no corresponding formulation for the scale-invariant condition used in this section.
The quantities used below, the incoherent conductivity $\sigma_Q$ and the Drude weights, are obtained from static horizon data and from thermodynamics, respectively, and do not rely on such a prescription.
The finite, universal DC quantity is the incoherent conductivity $\sigma_Q$~\cite{ref:DGH}.
For an Einstein--Maxwell--dilaton theory with the Maxwell normalization of~\eqref{eq:action}, it is given by the horizon formula~\cite{ref:DGH}
\begin{equation}
\sigma_Q = \frac{Z(\phi_h)}{4\pi G_4}\left(\frac{sT}{\sE+p}\right)^{2},
\qquad \sE+p = sT+\mu_Q\rho .
\label{eq:sigmaQ}
\end{equation}
At zero density it reduces to the membrane-paradigm result~\cite{ref:HKSS,ref:Iqbal-Liu,ref:Son-Starinets},
\begin{equation}
\sigma_Q^{(0)} = \frac{Z\bigl(\phi_0(r_h)\bigr)}{4\pi G_4} = \frac{e^{3\alpha/2}}{4\pi G_4},
\label{eq:sigma_DC}
\end{equation}
which, being dimensionless in $2+1$ dimensions, depends only on the marginal coupling and increases monotonically with $\alpha$.

\paragraph{Charge diffusion.}
With the susceptibility~\eqref{eq:chi}, the Einstein relation gives the charge diffusion constant~\cite{ref:DGH}
\begin{equation}
D_Q = \frac{\sigma_Q}{\chi} = l\,e^{3\alpha/2}\,\mu_Q^{(1)}(\alpha),\qquad
\frac{4\pi T\,D_Q}{3}=e^{2\alpha}\,j(\alpha),
\label{eq:DQ}
\end{equation}
which is independent of the normalization of the Maxwell term.
As required by conformal invariance, $TD_Q$ depends only on the coupling $\alpha$.
It equals $3/(4\pi)$ at $\alpha=0$, reproducing the $\AdS_4$-Schwarzschild value $D_Q=3/(4\pi T)$~\cite{ref:HerzogM,ref:KovtunStarinets}.
The combination $4\pi TD_Q/3$ decreases monotonically to $1/4$ as $\alpha\to\infty$, and by~\eqref{eq:j_small} its departure from $1$ at small $\alpha$ is non-analytic, $\propto\alpha^{1/3}$.

\paragraph{Charge correction to $\sigma_Q$.}
At order $\eps^2$ the incoherent conductivity is modified in two ways.
First, $Z$ is evaluated at the horizon value $\phi(r_*) = \phi_0(r_h) + \eps^2\bigl[\delta\phi(r_h) + \phi_0'(r_h)\,\delta r_h\bigr]$, with $\phi_0'(r_h)=3\sqrt7\,e^{-\alpha}\sinh\alpha/r_h$.
Second, the kinematic factor in~\eqref{eq:sigmaQ} deviates from unity, $\bigl(sT/(\sE+p)\bigr)^2 = 1 - 2\mu_Q\rho/(sT) + \order{\eps^4}$, with $\mu_Q\rho/(sT) = \eps^2\,4l\,\mu_Q^{(1)}/(3r_h^3)$. Altogether,
\begin{equation}
\frac{\delta\sigma_Q}{\sigma_Q^{(0)}} = \eps^2\left[\frac{3}{\sqrt7}\,\delta\phi(r_h) + 9\,e^{-\alpha}\sinh\alpha\,\frac{\delta r_h}{r_h} - \frac{8\,l\,\mu_Q^{(1)}}{3\,r_h^3}\right].
\label{eq:sigma_correction}
\end{equation}
In the labeling of Table~\ref{tab:bc}, $\delta r_h=0$, and~\eqref{eq:sigma_correction} is the correction at fixed $(S,\alpha)$.
At $\alpha=0.6$ the two remaining terms are $-0.0475$ and $-0.1682$, so that $\delta\sigma_Q/\sigma_Q^{(0)}\approx-0.216\,\eps^2$.
The correction is dominated by the kinematic term, and its magnitude decreases monotonically with $\alpha$ (Table~\ref{tab:transport_full}).

\subsection{Thermoelectric response}
\label{sec:transport:thermoelectric}

Two further combinations of~\eqref{eq:hydro_cond} remain meaningful in the clean limit.
The thermopower is the ratio of the Drude weights of $\vartheta$ and $\sigma$,
\begin{equation}
\Theta \equiv \lim_{\omega\to0}\frac{\vartheta}{\sigma} = \frac{s}{\rho} = \frac{\pi\, l\, r_h^2\,e^{\alpha/2}}{\eps},
\label{eq:Seebeck}
\end{equation}
which is independent of $\sigma_Q$. The thermal conductivity at zero electric current, $\kappa \equiv \bar\kappa - T\vartheta^2/\sigma$, is finite and controlled by $\sigma_Q$~\cite{ref:HLS,ref:DGH}:
\begin{equation}
\kappa = \frac{(\sE+p)^2\,\sigma_Q}{T\rho^2}
\simeq \frac{s^2\,T\,\sigma_Q^{(0)}}{\rho^2}
= \frac{e^{3\alpha/2}}{4\pi G_4}\,\frac{\pi^2 l^2 r_h^4\, e^{\alpha}\,T}{\eps^2},
\label{eq:kappa_bar}
\end{equation}
where $\simeq$ denotes the leading order in $\eps$.
Both quantities grow as $\eps\to0$, as $\eps^{-1}$ and $\eps^{-2}$ respectively.
This growth is not a clean-limit divergence---$\kappa$ is finite at any nonzero density---but reflects the fact that, as $\rho\to0$, the condition of vanishing electric current no longer constrains the momentum, which then carries the heat current.
Correspondingly, the ratio $\kappa/(T\sigma_Q)=(\sE+p)^2/(T^2\rho^2)\simeq s^2/\rho^2$ is parametrically large at small density, far from the Wiedemann--Franz value, while the conventional Lorenz ratio built from the (infinite) DC conductivity $\sigma$ vanishes in the clean limit.

\subsection{Hydrodynamic modes}
\label{sec:transport:modes}

At leading order in $\eps$, with $\sE+p=sT$, $\eta/s=1/(4\pi)$, $c_s^2=1/2$ and $\zeta=0$, the hydrodynamic poles at small wavevector $k$ are
\begin{align}
\omega_{\rm shear}  &= -\frac{i\,k^2}{4\pi T}+\order{k^4},
\label{eq:shear_mode}\\[2pt]
\omega_{\rm sound}  &= \pm\,\frac{k}{\sqrt2} - \frac{i\,k^2}{8\pi T} + \order{k^3},
\label{eq:sound_mode}\\[2pt]
\omega_{\rm charge} &= -i\,D_Q\,k^2 + \order{k^4}
 = -\frac{3i\,e^{2\alpha}j(\alpha)}{4\pi T}\,k^2+\order{k^4}.
\label{eq:charge_mode}
\end{align}
The shear and sound poles are independent of $\alpha$, while the charge diffusion pole depends on the marginal coupling through $e^{2\alpha}j(\alpha)$.
As stressed in Section~\ref{sec:transport:sigma}, these poles follow from hydrodynamics with the transport coefficients above.
The shear and charge-diffusion poles at zero density do not involve the scalar source; they are the same for both boundary conditions and are confirmed as quasinormal modes in Section~\ref{sec:dir:qnm}. The sound pole~\eqref{eq:sound_mode} is not: with the normalizable condition at large radius the bulk sound pole is that of the theory with fixed source, and realizing~\eqref{eq:mixedBC} at nonzero wavevector would require the growing asymptotic solution (Section~\ref{sec:dir:qnm}).

\subsection{Summary table}
\label{sec:transport:summary}

\begin{table}[ht]
\centering
\caption{Transport data in units $16\pi G_4 = l = m = 1$. Closed-form entries: $4\pi G_4\,\sigma_Q^{(0)}=e^{3\alpha/2}$ and $4\pi TD_Q/3=e^{2\alpha}j(\alpha)$; in addition $\eta/s=1/(4\pi)$, $c_s^2=1/2$ and $\zeta=0$ for all $\alpha$. The last column is the $\eps^2$ coefficient of~\eqref{eq:sigma_correction} at fixed $(S,\alpha)$.}
\label{tab:transport_full}
\renewcommand{\arraystretch}{1.15}
\begin{tabular}{cccc}
\toprule
$\alpha$ & $4\pi G_4\,\sigma_Q^{(0)}$ & $4\pi T D_Q/3$ & $\delta\sigma_Q/(\eps^2\sigma_Q^{(0)})$ \\
\midrule
0.2 & 1.3499 & 0.3790 & $-0.3975$ \\
0.3 & 1.5683 & 0.3410 & $-0.3311$ \\
0.4 & 1.8221 & 0.3173 & $-0.2831$ \\
0.5 & 2.1170 & 0.3012 & $-0.2458$ \\
0.6 & 2.4596 & 0.2897 & $-0.2157$ \\
0.7 & 2.8577 & 0.2812 & $-0.1906$ \\
0.8 & 3.3201 & 0.2747 & $-0.1693$ \\
1.0 & 4.4817 & 0.2658 & $-0.1350$ \\
1.5 & 9.4877 & 0.2555 & $-0.0794$ \\
2.0 & 20.086 & 0.2520 & $-0.0476$ \\
\bottomrule
\end{tabular}
\end{table}

\section{Second-order observables and their verification}
\label{sec:second}

The preceding sections determine the complete $\order{\eps^2}$ solution at fixed $(S,\alpha)$.
In this section we collect the resulting observables, derive one further quantity that follows from the second-order fields, namely the third-order chemical potential, which fixes the quartic term of the pressure, and verify all second-order results by independent methods.

\subsection{Summary of the second-order data}
\label{sec:second:data}

At fixed entropy and fixed $\alpha$ the second-order corrections are
\begin{equation}
\frac{T^{(2)}}{T^{(0)}}=-\frac{\mu_Q^{(1)}}{6ml},\qquad
\frac{S^{(2)}}{S^{(0)}}\bigg|_{T}=\frac{\mu_Q^{(1)}}{3ml},\qquad
\sE^{(2)}=\frac{\mu_Q^{(1)}}{8\pi G_4 l},\qquad
p^{(2)}=\tfrac12\sE^{(2)},
\label{eq:second_summary}
\end{equation}
together with the source shift $\delta\varphi_{(0)}$ of~\eqref{eq:dphi0}, the condensate shift $\delta\langle\mathcal O_\phi\rangle/\langle\mathcal O_\phi\rangle=-\tfrac{2}{\sqrt7}\,\delta\varphi_{(0)}$ implied by~\eqref{eq:mixedBC}, the horizon scalar $\delta\phi(r_h)$ and the conductivity correction~\eqref{eq:sigma_correction}.
The first four follow in closed form from the first law and the susceptibility; the remaining ones require the second-order solution.
Table~\ref{tab:thermo} lists all of them, together with the quartic coefficient derived next.

\subsection{Third-order chemical potential and the quartic pressure}
\label{sec:second:mu3}

\begin{figure}[ht]
\centering
\includegraphics[width=0.98\textwidth]{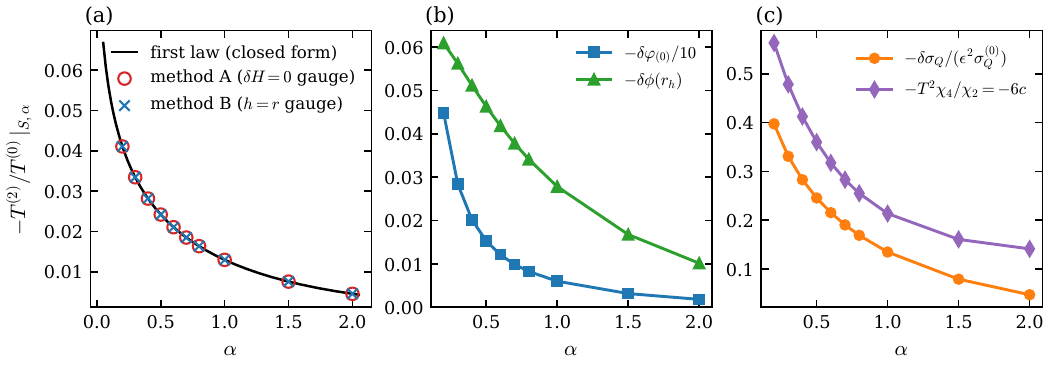}
\caption{Second-order observables at fixed $(S,\alpha)$. (a)~Temperature correction: the closed form~\eqref{eq:T2_closed} (solid) and the two independent perturbative computations, methods A and B of Section~\ref{sec:second:checks} (symbols).
(b)~Source shift~\eqref{eq:dphi0} (scaled by $1/10$) and horizon value of the scalar perturbation.
(c)~Charge correction to the incoherent conductivity, Eq.~\eqref{eq:sigma_correction}, and the cumulant ratio~\eqref{eq:chi4}. Since all these quantities are negative, their magnitudes are plotted. All charge corrections are $\eps^2$ coefficients in units $16\pi G_4=l=m=1$.}
\label{ChargedScalarfig}
\end{figure}

At fixed charge density the chemical potential is odd in $\eps$, $\mu_Q=\eps\,\mu_Q^{(1)}+\eps^3\mu_Q^{(3)}+\order{\eps^5}$.
Its third-order coefficient requires only the second-order fields. By the Gauss law $A_t'=Q\,ab/(h^2Z)$, and in the gauge $\delta H\equiv0$ one has $ab=a_0b_0(1+\tfrac12\eps^2g)$ and $Z=Z(\phi_0)\bigl(1+\tfrac{3}{\sqrt7}\eps^2\delta\phi\bigr)$.
For the solution with unshifted source ($\delta r_h=0$, $\delta\phi_\infty=\delta A_\infty=0$) this gives
\begin{equation}
\mu_Q^{(3)}\big|_{S,\src}=\int_{r_h}^{\infty}A_t^{(1)\prime}(r)\Bigl[\frac{g(r)}{2}-\frac{3}{\sqrt7}\,\delta\phi(r)\Bigr]dr ,
\label{eq:mu3}
\end{equation}
and the relabeling to fixed $\alpha$ adds $(\partial\mu_Q^{(1)}/\partial\varphi_{(0)})|_S\,\delta\varphi_{(0)}$.
Expanding $\rho/\mu_Q$ at fixed $(S,\alpha)$ and using conformal invariance, we obtain
\begin{equation}
\frac{\rho}{\mu_Q}=\chi^{(0)}\Bigl[1+c\,\frac{\mu_Q^2}{T^2}+\order{\mu_Q^4}\Bigr],\qquad
c(\alpha)=-\Bigl(\frac{T^{(2)}}{T^{(0)}}+\frac{\mu_Q^{(3)}}{\mu_Q^{(1)}}\Bigr)\Bigl(\frac{T^{(0)}}{\mu_Q^{(1)}}\Bigr)^2 ,
\label{eq:c_def}
\end{equation}
where all quantities are evaluated at fixed $(S,\alpha)$.
The pressure and the susceptibility at fixed temperature follow as
\begin{equation}
p=p^{(0)}(T)+\tfrac12\chi^{(0)}\mu_Q^2+\tfrac14\,c\,\chi^{(0)}\frac{\mu_Q^4}{T^2}+\order{\mu_Q^6},\qquad
\chi=\chi^{(0)}\Bigl(1+3c\,\frac{\mu_Q^2}{T^2}\Bigr),\qquad
\frac{T^2\chi_4}{\chi_2}=6c ,
\label{eq:chi4}
\end{equation}
where $\chi_n\equiv\partial^n p/\partial\mu_Q^n|_{\mu_Q=0}$. Since $\chi^{(0)}\propto T$, the ratio $T^2\chi_4/\chi_2$ is a dimensionless function of the coupling $\alpha$ only, the analogue of the cumulant ratios used to characterize finite-density matter.
We find $c<0$ for all ten values of $\alpha$ considered, with $|c|$ decreasing from $0.094$ at $\alpha=0.2$ to $0.024$ at $\alpha=2.0$ (Table~\ref{tab:thermo} and Figure~\ref{ChargedScalarfig}c).
The susceptibility therefore initially decreases with the chemical potential, in quantitative agreement with the nonlinear solutions of Section~\ref{sec:finite}, where $\chi/\chi^{(0)}\simeq1-0.159\,(\mu_Q/T)^2$ at small $\mu_Q/T$ for $\alpha=0.6$ and reaches a minimum of $0.960$ at $\mu_Q/T\approx0.81$.

\begin{table}[ht]
\centering
\caption{Complete second-order data at fixed $(S,\alpha)$ in units $16\pi G_4=l=m=1$ (so $r_h=2^{1/3}$). Columns 3--6 are $\eps^2$ coefficients: temperature shift~\eqref{eq:T2_closed}, source shift~\eqref{eq:dphi0}, horizon scalar and conductivity correction~\eqref{eq:sigma_correction}. The last two columns give the quartic coefficient~\eqref{eq:c_def} and the cumulant ratio~\eqref{eq:chi4}. All entries agree between the independent methods of Section~\ref{sec:second:checks} to the accuracy given in Table~\ref{tab:second_checks}.}
\label{tab:thermo}
\small
\setlength{\tabcolsep}{4pt}
\renewcommand{\arraystretch}{1.15}
\begin{tabular}{cccccccc}
\toprule
$\alpha$ & $\mu_Q^{(1)}$ & $T^{(2)}/T^{(0)}$ & $\delta\varphi_{(0)}$ & $\delta\phi(r_h)$ & $\delta\sigma_Q/\sigma_Q^{(0)}$ & $c$ & $T^2\chi_4/\chi_2$ \\
\midrule
0.2 & 0.246303 & $-0.0410505$ & $-0.4481$ & $-0.06090$ & $-0.3975$ & $-0.09387$ & $-0.5632$ \\
0.3 & 0.200485 & $-0.0334142$ & $-0.2839$ & $-0.05622$ & $-0.3311$ & $-0.07973$ & $-0.4784$ \\
0.4 & 0.168808 & $-0.0281347$ & $-0.2017$ & $-0.05116$ & $-0.2831$ & $-0.06875$ & $-0.4125$ \\
0.5 & 0.144998 & $-0.0241663$ & $-0.1531$ & $-0.04631$ & $-0.2458$ & $-0.05999$ & $-0.3600$ \\
0.6 & 0.126182 & $-0.0210304$ & $-0.1212$ & $-0.04185$ & $-0.2157$ & $-0.05293$ & $-0.3176$ \\
0.7 & 0.110813 & $-0.0184688$ & $-0.0989$ & $-0.03779$ & $-0.1906$ & $-0.04719$ & $-0.2832$ \\
0.8 & 0.097961 & $-0.0163269$ & $-0.0826$ & $-0.03413$ & $-0.1693$ & $-0.04253$ & $-0.2552$ \\
1.0 & 0.077610 & $-0.0129350$ & $-0.0602$ & $-0.02784$ & $-0.1350$ & $-0.03563$ & $-0.2138$ \\
1.5 & 0.045249 & $-0.0075414$ & $-0.0317$ & $-0.01679$ & $-0.0794$ & $-0.02682$ & $-0.1609$ \\
2.0 & 0.027067 & $-0.0045112$ & $-0.0183$ & $-0.01016$ & $-0.0476$ & $-0.02359$ & $-0.1415$ \\
\bottomrule
\end{tabular}
\end{table}

\subsection{Independent verification}
\label{sec:second:checks}

Each second-order quantity is computed by the following independent methods.
\begin{itemize}
\item[(A)] Perturbation theory in the gauge $\delta H\equiv0$ (Sections~\ref{sec:secondorder}--\ref{sec:horizon}). The solution with unshifted source is constructed numerically and then relabeled to fixed $\alpha$ through the Dirichlet first law, Eq.~\eqref{eq:dphi0}.
\item[(B)] Perturbation theory in the gauge $h=r$ of Section~\ref{sec:finite}. The full equations~\eqref{eq:full_F}--\eqref{eq:full_phi} are linearized in $Q^2$ around the neutral branes. The horizon radius is held fixed, so the entropy is fixed automatically, and the boundary condition is imposed directly by requiring the Ward-identity coupling~\eqref{eq:full_k} to be unchanged. The energy is computed from the Brown--York data~\eqref{eq:full_EP} with $A_3$ fitted from the asymptotic tail rather than taken from the Noether identity. This method shares neither the gauge, nor the code, nor the implementation of the boundary condition with method~A.
\item[(C)] Fully nonlinear solutions (Section~\ref{sec:finite}) at small charge, $0.03\le Q\le0.16$, from which the $\order{Q^2}$ coefficients are extracted by polynomial extrapolation in $Q^2$ (for $\alpha=0.3$, $0.6$, $1.0$ and $1.5$).
\item[(D)] The closed forms~\eqref{eq:second_summary}, which follow from the first law and the closed-form susceptibility alone. They exist for $T^{(2)}$, $S^{(2)}$, $\sE^{(2)}$ and $p^{(2)}$ only; for $\delta\phi(r_h)$, $\delta\sigma_Q$ and $c$ there is no closed form, and $\delta\varphi_{(0)}$ enters method A through~\eqref{eq:dphi0} and is therefore tested by the comparison with method B.
\end{itemize}
To compare methods A and C with method B, the latter's coefficients are converted using the scale-invariant ratio $\rho/s$.
Table~\ref{tab:second_checks} lists the largest relative deviations.
Methods A and B agree to better than $1.3\times10^{-6}$ for all quantities and all ten values of $\alpha$; where a closed form exists, method D agrees with them to $4\times10^{-7}$ for $T^{(2)}$ and to $4\times10^{-6}$ for $\sE^{(2)}$.
Method C, which is limited by the precision of extracting $\order{Q^2}$ and $\order{Q^4}$ coefficients from nonlinear data, agrees to $3\times10^{-6}$--$1.2\times10^{-5}$ for the second-order quantities and to $6\times10^{-4}$ for $c$, a third-order quantity.
In addition, the solution with unshifted source satisfies the Maxwell relation of the Dirichlet ensemble to $10^{-8}$ (Section~\ref{sec:thermo:check}).

\begin{table}[ht]
\centering
\caption{Independent verification of the second-order results: largest relative deviation between methods over all values of $\alpha$ considered (ten values for A, B and D; four for C). The second column states which methods are compared; the closed forms (D) exist only for $T^{(2)}$ and $\sE^{(2)}$. Relative deviations are independent of units.}
\label{tab:second_checks}
\small
\renewcommand{\arraystretch}{1.15}
\begin{tabular}{llcc}
\toprule
Quantity & Methods compared & Max.\ deviation & C vs.\ B \\
\midrule
$T^{(2)}/T^{(0)}$ & A vs.\ B vs.\ D & $4\times10^{-7}$ & $3\times10^{-6}$ \\
$\sE^{(2)}$ & B (direct Brown--York) vs.\ D & $4\times10^{-6}$ & -- \\
$\delta\varphi_{(0)}$ & A (via Eq.~\eqref{eq:dphi0}) vs.\ B & $3\times10^{-7}$ & $3\times10^{-6}$ \\
$\delta\phi(r_h)$ & A vs.\ B & $4\times10^{-7}$ & $1.2\times10^{-5}$ \\
$\delta\sigma_Q/\sigma_Q^{(0)}$ & A vs.\ B & $1\times10^{-7}$ & $3\times10^{-6}$ \\
$c$ (from $\mu_Q^{(3)}$) & A vs.\ B & $1.3\times10^{-6}$ & $6\times10^{-4}$ \\
\bottomrule
\end{tabular}
\end{table}

\section{Finite charge density: nonlinear solutions}
\label{sec:finite}

The perturbative analysis of the preceding sections is valid for $\mu_Q/T\ll1$, and the nonlinear solutions constructed here provide method C of Section~\ref{sec:second:checks}.
In this section we construct the charged black branes at arbitrary charge density by solving the full nonlinear field equations numerically, compute their thermodynamic and transport properties, and verify them against the exact relations and the perturbative results.
We set $l=1$ and quote numerical values in units $16\pi G_4=1$.

\subsection{Field equations}
\label{sec:finite:eqs}

It is convenient to use the transverse warp factor as the radial coordinate,
\begin{equation}
ds^2=-F(r)\,e^{-\psi(r)}dt^2+\frac{dr^2}{F(r)}+r^2\,(dx^2+dy^2),
\label{eq:hgauge}
\end{equation}
together with $\phi(r)$ and the gauge field fixed by the Gauss law, $A_t'=Q\,e^{-\psi/2}/(r^2Z(\phi))$. The $tt$ and $rr$ Einstein equations and the scalar equation reduce to
\begin{align}
F' &= -F\Bigl(\frac{r\phi'^2}{4}+\frac1r\Bigr)-\frac{Q^2e^{-3\phi/\sqrt7}}{r^3}+\frac{21r}{4}\,e^{-3\phi/\sqrt7}-\frac{9r}{4}\,e^{-\sqrt7\phi},
\label{eq:full_F}\\
\psi' &= -\frac{r\,\phi'^2}{2},
\label{eq:full_chi}\\
F r^4 e^{\sqrt7\phi}\,\phi'' &= -F r^3 e^{\sqrt7\phi}\phi'+e^{4\phi/\sqrt7}\Bigl[Q^2 r\,\phi'-\tfrac{6\sqrt7}{7}Q^2-\tfrac{21}{4}r^5\phi'+\tfrac{9\sqrt7}{2}r^4\Bigr]+\tfrac94 r^5\phi'-\tfrac{9\sqrt7}{2}r^4 ,
\label{eq:full_phi}
\end{align}
and the $xx$ equation is then satisfied identically. We have verified this symbolically, and along the numerical solutions to a relative accuracy of $10^{-6}$ or better.
At a regular horizon $F(r_h)=0$, and the right-hand side of~\eqref{eq:full_phi} must vanish there, which fixes
\begin{equation}
\phi'(r_h)=\frac{\tfrac{6\sqrt7}{7}Q^2e^{4\phi_h/\sqrt7}-\tfrac{9\sqrt7}{2}r_h^4\bigl(e^{4\phi_h/\sqrt7}-1\bigr)}{Q^2r_h\,e^{4\phi_h/\sqrt7}-\tfrac{21}{4}r_h^5e^{4\phi_h/\sqrt7}+\tfrac94r_h^5}.
\label{eq:full_bc}
\end{equation}
Using the scaling symmetry we set $r_h=1$, so that a solution is labeled by the horizon value $\phi_h$ and the charge $Q$; the value $\psi(r_h)$ is a time normalization, fixed below.
The entropy and charge densities are $s=r_h^2/(4G_4)$ and $\rho=Q/(4\pi G_4)$.
With $N^2\equiv\lim_{r\to\infty}F e^{-\psi}/r^2$, the boundary time normalization $a^2/h^2\to1$ gives
\begin{equation}
T=\frac{F'(r_h)\,e^{-\psi(r_h)/2}}{4\pi N},\qquad
\mu_Q=\frac1N\int_{r_h}^\infty\frac{Q\,e^{-\psi/2}}{r^2Z(\phi)}\,dr .
\label{eq:full_Tmu}
\end{equation}

\subsection{Holographic data at finite charge}
\label{sec:finite:holo}

To read off the renormalized stress tensor we use the scalar itself to define a radial coordinate $\tilde r$ through $e^{2\phi/\sqrt7}=\src\,\tilde r^3$, where $\src$ is the invariant source~\eqref{eq:src_inv}.
For every solution of~\eqref{eq:full_F}--\eqref{eq:full_phi},
\begin{equation}
a^2=\src^{1/2}\tilde r^{7/2}\Bigl(1+\frac{A_3}{\tilde r^3}+\cdots\Bigr),\qquad
h^2=\src^{1/2}\tilde r^{7/2}\Bigl(1+\frac{H_3}{\tilde r^3}+\cdots\Bigr),
\label{eq:full_asym}
\end{equation}
in boundary-normalized time, where the omitted terms are of order $\tilde r^{-6}$ and the Maxwell field first contributes at relative order $Q^2\tilde r^{-7}/\src$.
Since the homogeneous modes behave as $\tilde r^{0}$ and $\tilde r^{-3}$, the Maxwell source is non-resonant and induces no logarithms.
Inserting~\eqref{eq:full_asym} into~\eqref{eq:BYmixed}, with $b^2$ determined by the $rr$ constraint, we find that the divergences cancel and
\begin{equation}
\sE_{\rm D}=\frac{1}{16\pi G_4}\Bigl(-\tfrac72A_3-H_3-\frac3\src\Bigr),\qquad
p_{\rm D}=\frac{1}{16\pi G_4}\Bigl(\tfrac12A_3+4H_3+\frac3\src\Bigr).
\label{eq:full_EP}
\end{equation}
For the neutral branes these expressions reproduce~\eqref{eq:eps_p}.
The field equations imply that
\begin{equation}
\Xi(r)\equiv\frac{1}{16\pi G_4}\,e^{\psi/2}r^4\Bigl(\frac{Fe^{-\psi}}{r^2}\Bigr)'-\rho\,A_t(r)
\label{eq:full_N}
\end{equation}
is independent of $r$ (with $A_t(r_h)=0$ and $t$ the integration time).
Equating its horizon value, $N\,Ts$, to its boundary value, $N(\sE_{\rm D}+p_{\rm D}-\mu_Q\rho)$, gives the Smarr-type identity
\begin{equation}
\sE_{\rm D}+p_{\rm D}=\frac{3(H_3-A_3)}{16\pi G_4}=Ts+\mu_Q\rho .
\label{eq:full_noether}
\end{equation}
The condensate follows from the trace Ward identity, $\langle\mathcal O_\phi\rangle=(-\sE_{\rm D}+2p_{\rm D})/\varphi_{(1)}$, and the scale-invariant boundary condition~\eqref{eq:mixedBC} requires the coupling
\begin{equation}
k=\frac{\src}{3}\bigl(-\sE_{\rm D}+2p_{\rm D}\bigr)
\label{eq:full_k}
\end{equation}
to take the value~\eqref{eq:k_alpha} of the chosen $\alpha$.
In this ensemble $\sE=\sE_{\rm D}+\mathcal W=\tfrac23(Ts+\mu_Q\rho)$ and $p=\sE/2$.

\subsection{Numerical construction}
\label{sec:finite:num}

We integrate~\eqref{eq:full_F}--\eqref{eq:full_phi} outward in $\ln r$ from $r_h(1+10^{-10})$, starting from the regular data~\eqref{eq:full_bc}, up to $r=10^5$, using an eighth-order Runge--Kutta method (DOP853) with relative tolerance $10^{-13}$.
The chemical potential is integrated along with the solution.
From the asymptotic region we extract $\src$ from $(2/\sqrt7)\phi-(12/7)\ln r$, $H_3$ from the tail of $r\phi'$, and $N$ from $Fe^{-\psi}/r^2$, each by a least-squares fit in powers of $\tilde r^{-3}$ over $300\le r\le3\times10^4$. For $\src$ and $N$ the fit also includes the Maxwell-induced powers $\tilde r^{-7}$ and $\tilde r^{-10}$; for $H_3$ their effect is below $10^{-9}$.
We then determine $A_3$ from the Noether identity~\eqref{eq:full_noether}, and use an independent fit of $A_3$ from the tail of $(\ln a^2)'$ as a check.
For a given $\alpha$ and charge $Q$, we solve~\eqref{eq:full_k} for $\phi_h$ by bracketing and Brent's method, and continue in $Q$ from the neutral brane toward the extremal limit.
We constructed about $45$ solutions for each of $\alpha=0.3$, $0.6$, $1.0$ and $1.5$, covering $0\le\mu_Q/T\lesssim150$--$240$, i.e.\ down to $T/\mu_Q\approx0.004$, and continued the $\alpha=0.6$ branch to $T/\mu_Q=3\times10^{-4}$.

\subsection{Thermodynamics}
\label{sec:finite:thermo}

Since the equilibrium thermodynamics is conformal at fixed $\alpha$, it is encoded in a single function of $\mu_Q/T$,
\begin{equation}
p(T,\mu_Q;\alpha)=T^3\,\Pi\bigl(\mu_Q/T;\alpha\bigr),\qquad
\rho=T^2\Pi',\qquad s=T^2\bigl(3\Pi-x\Pi'\bigr),\qquad \sE=2p,\qquad x\equiv\mu_Q/T,
\label{eq:full_g}
\end{equation}
and the charged Euler relation~\eqref{eq:euler} holds at all densities.
Table~\ref{tab:finite} lists, for $\alpha=0.6$, the entropy, pressure, charge susceptibility $\chi=\partial\rho/\partial\mu_Q|_T$ and incoherent conductivity, normalized to their neutral values at the same temperature, together with the horizon scalar.
Table~\ref{tab:finite2} compares different values of $\alpha$.

\begin{figure}[ht]
\centering
\includegraphics[width=0.98\textwidth]{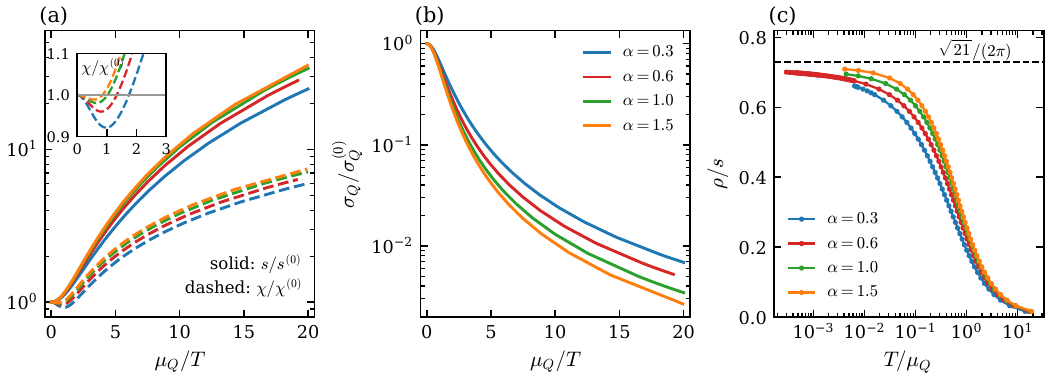}
\caption{Nonlinear charged branes at fixed $\alpha$. (a)~Entropy density (solid) and charge susceptibility (dashed) relative to their neutral values at the same temperature; the inset shows the initial decrease of the susceptibility, predicted by $c<0$ in~\eqref{eq:chi4}. (b)~Incoherent conductivity relative to its neutral value. (c)~The ratio $\rho/s$ as a function of $T/\mu_Q$; the dashed line is the bound~\eqref{eq:rho_s_bound}, which is saturated in the extremal limit, Eq.~\eqref{eq:saturation}.}
\label{fig:finite}
\end{figure}

Several features stand out. First, $s/s^{(0)}$ and $p/p^{(0)}$ grow monotonically with $\mu_Q/T$, and more rapidly for larger $\alpha$ (Figure~\ref{fig:finite}a). Second, the susceptibility is non-monotonic (inset of Figure~\ref{fig:finite}a): $\chi/\chi^{(0)}$ first decreases, reaching a minimum ($0.921$, $0.960$, $0.981$ and $0.989$ at $\mu_Q/T\approx0.98$, $0.81$, $0.66$ and $0.56$ for $\alpha=0.3$, $0.6$, $1.0$ and $1.5$), and then grows roughly linearly at large $\mu_Q/T$.
Third, the specific heats at fixed chemical potential and at fixed charge, the susceptibility and the determinant of the Hessian of $p(T,\mu_Q)$ are positive for all solutions we constructed (we checked this with interpolated derivatives over the full data sets, including the $\alpha=0.6$ branch down to $T/\mu_Q=3\times10^{-4}$, and with local finite differences up to $\mu_Q/T\approx146$), so these branes are locally thermodynamically stable in both the grand-canonical and canonical ensembles.
(Global stability is a separate issue, discussed in Section~\ref{sec:thermo:stability}. Dynamical stability at nonzero wavevector is also separate: the neutral branes with $\alpha>\tfrac12\ln5$, from which the branches with $\alpha=1.0$ and $1.5$ start, have a growing sound mode (Section~\ref{sec:dir:qnm}); we have not studied the charged branes of these branches dynamically.)
Fourth, the condensate $\langle\mathcal O_\phi\rangle$ is negative throughout and grows in magnitude with the density.

\paragraph{Extremal limit.}
The low-temperature behavior can be understood analytically. At the horizon,~\eqref{eq:full_F} gives
\begin{equation}
F'(r_h)=e^{-3\phi_h/\sqrt7}\Bigl(\frac{21}{4}r_h-\frac{Q^2}{r_h^3}\Bigr)-\frac{9}{4}\,r_h\,e^{-\sqrt7\phi_h}.
\label{eq:Fprime_h}
\end{equation}
Because $Z$ and the leading term of $V$ carry opposite exponentials, the Maxwell term and the leading potential term have the same $\phi_h$ dependence.
Non-negativity of $T$ then requires
\begin{equation}
Q^2\le\frac{r_h^4}{4}\bigl(21-9\,e^{-4\phi_h/\sqrt7}\bigr)<\frac{21}{4}\,r_h^4
\qquad\Longrightarrow\qquad
\frac{\rho}{s}<\frac{\sqrt{21}}{2\pi},
\label{eq:rho_s_bound}
\end{equation}
an exact bound valid for all $\alpha$ and $T$ (with $l=1$).
Furthermore, a regular extremal horizon cannot exist at finite $\phi_h$. When $F'(r_h)=0$, the denominator of~\eqref{eq:full_bc} vanishes identically, while its numerator reduces to $(18\sqrt7/7)\,r_h^4\neq0$.
The scalar equation therefore has no solution that is regular at a double zero of $F$, and as $T\to0$ the horizon value $\phi_h$ must run to infinity.
On the other hand, the bound~\eqref{eq:rho_s_bound} holds at every $T>0$, so that $s\ge2\pi\rho/\sqrt{21}$: at fixed charge density the entropy density remains finite as $T\to0$.

\paragraph{Infrared throat.}
These two facts determine the zero-temperature geometry. Near the horizon the transverse warp factor tends to a constant $h_0$, and the scalar moves in the effective potential
\begin{equation}
V_{\rm eff}(\phi)=V(\phi)+\frac{2Q^2}{Z(\phi)\,h_0^4}
=-2\varepsilon\,e^{-3\phi/\sqrt7}+\frac92\,e^{-\sqrt7\phi},
\qquad
\varepsilon\equiv\frac{21}{4}-\frac{Q^2}{h_0^4},
\label{eq:Veff_IR}
\end{equation}
where $2Q^2/(Zh_0^4)$ is the electric contribution in our normalization of the Maxwell term. (This bulk effective potential should not be confused with the boundary potential $\mathcal V$ of Section~\ref{sec:thermo:stability}.)
As noted above, the Maxwell contribution has the same $\phi$ dependence as the leading term of $V$. For $\varepsilon=0$, and without the second exponential of $V$, every constant value of $\phi$ gives an exact $\AdS_2\times\mathbb R^2$ solution: the Maxwell field and the leading potential create a flat direction.
The second exponential lifts this degeneracy and drives $\phi$ slowly to infinity. The resulting geometry is described by the following proposition, which we derive in Appendix~\ref{app:throat}.

\begin{proposition}\label{prop:throat}
Consider a zero-temperature solution whose near-horizon region, in the coordinates $ds^2=-D\,dt^2+du^2/D+H^2(dx^2+dy^2)$ with the horizon at $u\to0$, has $H\to h_0>0$ and $\phi\to\infty$, with all fields varying slowly in $\tau\equiv\ln(1/u)$. Then $Q^2=\tfrac{21}{4}h_0^4$, i.e.\ $\varepsilon=0$, and for $\tau\to\infty$
\begin{equation}
e^{-4\phi/\sqrt7}\simeq\frac{49}{48\,\tau},\qquad
H\simeq h_0\Bigl(1+\frac{7}{64\,\tau}\Bigr),\qquad
D\simeq\frac{21}{2}\,e^{-3\phi/\sqrt7}\,u^2,
\label{eq:throat}
\end{equation}
with corrections of relative order $\ln\tau/\tau$. The solution depends on $h_0$ and on one integration constant, which shifts $\tau$, and it is independent of $\alpha$.
\end{proposition}

The throat is thus locally $\AdS_2\times\mathbb R^2$, with an $\AdS_2$ radius $L_2^2=\tfrac{2}{21}e^{3\phi/\sqrt7}\propto\tau^{3/4}$ that grows logarithmically, and the scalar runs as $\phi\simeq\tfrac{\sqrt7}{4}\ln\tau=\tfrac{\sqrt7}{4}\ln\ln(1/u)$, i.e.\ logarithmically in the $\AdS_2$ radial variable $\tau$. The flat direction is approached only as $\phi\to\infty$, so the infrared fixed point is reached logarithmically slowly, and the effective Maxwell coupling $1/Z(\phi)$ flows to zero.
The expansion in $1/\tau$ can be continued. At next order the relative correction to $e^{-3\phi/\sqrt7}$ contains a term $\tfrac{27}{64}\,\tau^{-1}\ln\tau$, and all coefficients are fixed except the shift of $\tau$ (Appendix~\ref{app:throat}).

The hypotheses of the proposition are forced by the results above, except for the slow variation: $\phi_h\to\infty$ because no regular extremal horizon exists at finite $\phi_h$, and $h_0>0$ because of the bound.
Within slowly varying throats, $\varepsilon\neq0$ is excluded. For $\varepsilon>0$ the scalar equation drives $\phi$ to decrease toward the horizon, and for $\varepsilon<0$ it drives $\phi$ to grow linearly in $\tau$, which forces $H\to0$ and contradicts $s\ge2\pi\rho/\sqrt{21}$ (Appendix~\ref{app:throat}).
With $s\to h_0^2/(4G_4)$ and $\rho=Q/(4\pi G_4)$, the condition $\varepsilon=0$ gives
\begin{equation}
\frac\rho s\;\longrightarrow\;\frac{\sqrt{21}}{2\pi}\approx0.7293,\qquad s\;\longrightarrow\;s_0\equiv\frac{2\pi\rho}{\sqrt{21}}\qquad(T\to0),
\label{eq:saturation}
\end{equation}
independently of $\alpha$, while $Z(\phi_h)$ diverges. The bound~\eqref{eq:rho_s_bound} is therefore saturated in the extremal limit.
We have not proven that a complete zero-temperature solution with this throat exists for every $\alpha$ and $\rho$; the evidence is the consistency of the $1/\tau$ expansion through next-to-leading order and the agreement with the numerical solutions described below.

\paragraph{Low-temperature behavior.}
At small $T$ the horizon lies at a large value $\tau_h$ in the throat. Over a range $\Delta\tau=\order{1}$ around $\tau_h$ the fields are nearly constant, and the geometry there is the black hole in $\AdS_2$ of radius $L_2(\tau_h)$. Hence $T\propto e^{-3\phi_h/\sqrt7}\,e^{-\tau_h}$, with a $T$-independent proportionality constant fixed by the ultraviolet region, and
\begin{equation}
\ln\frac{\mu_Q}{T}=\tau_h+\frac34\ln\tau_h+\order{1}.
\label{eq:tau_h}
\end{equation}
The horizon values follow from~\eqref{eq:throat} at $\tau=\tau_h$, with corrections of relative order $\ln\tau_h/\tau_h$:
\begin{gather}
\phi_h\simeq\frac{\sqrt7}{4}\ln\frac{48\,\tau_h}{49},\qquad
\frac{s}{s_0}\simeq1+\frac{7}{32\,\tau_h},\nonumber\\
\frac{\rho}{s}\simeq\frac{\sqrt{21}}{2\pi}\Bigl(1-\frac{7}{32\,\tau_h}\Bigr),\qquad
C_\rho\equiv T\Bigl(\frac{\partial s}{\partial T}\Bigr)_{\!\rho}\simeq\frac{7\,s_0}{32\,\tau_h^2}.
\label{eq:lowT}
\end{gather}
The horizon scalar therefore grows as $\ln\ln(\mu_Q/T)$, and the entropy and $\rho/s$ approach their extremal values only as $1/\ln(\mu_Q/T)$. The specific heat is positive, in agreement with the local stability found numerically, and vanishes as $1/\ln^2(\mu_Q/T)$ rather than linearly in $T$ as for an extremal Reissner--Nordstr\"om horizon.
The ratio $\rho/s$ in~\eqref{eq:lowT} coincides at this order with the horizon value of the bound~\eqref{eq:rho_s_bound}, $\sqrt{21-9e^{-4\phi_h/\sqrt7}}/(2\pi)$: the bracket in~\eqref{eq:Fprime_h} vanishes at leading order, and $T\to0$ is reached through the throat rather than through the prefactor $e^{-3\phi_h/\sqrt7}$.

\paragraph{Comparison with the numerical solutions.}
For $\alpha=0.6$ we followed the branch down to $T/\mu_Q=3\times10^{-4}$, where $\rho/s=0.701$ and $\phi_h=1.21$ (Figure~\ref{fig:finite}c). At this value of $\phi_h$ the horizon value of the bound is $0.704$, so the bracket in~\eqref{eq:Fprime_h} is already close to zero, as predicted. The bound~\eqref{eq:rho_s_bound} is satisfied by all solutions.
Equations~\eqref{eq:throat} and~\eqref{eq:tau_h} predict the local exponent
\begin{equation}
\frac{d\ln e^{-4\phi_h/\sqrt7}}{d\ln T}\simeq\frac{1}{\tau_h+3/4},\qquad \tau_h\simeq\frac{49}{48}\,e^{4\phi_h/\sqrt7}.
\label{eq:local_exp}
\end{equation}
The measured exponent is $0.17$ at $T/\mu_Q\approx5\times10^{-3}$ and $0.13$ at $3\times10^{-4}$. With the corresponding values $\phi_h\approx0.93$ and $1.21$ ($\tau_h\approx4.2$ and $6.4$), Eq.~\eqref{eq:local_exp} gives $0.20$ and $0.14$. The ratio of measured to predicted values grows from $0.84$ to $0.92$ as $\tau_h$ increases, as expected from corrections of relative order $\ln\tau_h/\tau_h$, which are still sizable in this range.
The slow drift of $s/\mu_Q^2$, $\rho/\mu_Q^2$ and $p/\mu_Q^3$ at low temperature (e.g.\ $s/\mu_Q^2=33.7$ at $T/\mu_Q=0.01$ and $31.2$ at $3\times10^{-4}$ for $\alpha=0.6$) is the logarithmic approach described by~\eqref{eq:lowT}.

\begin{table}[ht]
\centering
\caption{Finite-density thermodynamics and conductivity at $\alpha=0.6$, relative to the neutral values at the same temperature: $s^{(0)}\propto T^2$, $p^{(0)}=\tfrac13Ts^{(0)}$, $\chi^{(0)}$ from~\eqref{eq:chi} and $\sigma_Q^{(0)}$ from~\eqref{eq:sigma_DC}. The neutral horizon scalar is $\phi_h=0.7937$. All entries are dimensionless (units $16\pi G_4=l=1$ as in Section~\ref{sec:finite}).}
\label{tab:finite}
\renewcommand{\arraystretch}{1.15}
\begin{tabular}{cccccc}
\toprule
$\mu_Q/T$ & $s/s^{(0)}$ & $p/p^{(0)}$ & $\chi/\chi^{(0)}$ & $\sigma_Q/\sigma_Q^{(0)}$ & $\phi_h$ \\
\midrule
0.5 & 1.0329 & 1.0978 & 0.9724 & 0.8566 & 0.7646 \\
1   & 1.1312 & 1.3873 & 0.9654 & 0.6016 & 0.7056 \\
2   & 1.4965 & 2.5514 & 1.1334 & 0.2805 & 0.6135 \\
5   & 3.5335 & 12.285 & 1.9993 & 0.0631 & 0.5551 \\
10  & 9.4680 & 63.212 & 3.5291 & 0.0177 & 0.5867 \\
20  & 30.368 & 403.60 & 6.6018 & 0.0048 & 0.6554 \\
\bottomrule
\end{tabular}
\end{table}

\begin{table}[ht]
\centering
\caption{Dependence on the coupling $\alpha$: normalized quantities at $\mu_Q/T=5$, and quantities in units of $\mu_Q$ at $T/\mu_Q=0.01$ (units $16\pi G_4=l=1$).}
\label{tab:finite2}
\small
\setlength{\tabcolsep}{4pt}
\renewcommand{\arraystretch}{1.15}
\begin{tabular}{c|ccccc|cccc}
\toprule
 & \multicolumn{5}{c|}{$\mu_Q/T=5$} & \multicolumn{4}{c}{$T/\mu_Q=0.01$} \\
$\alpha$ & $s/s^{(0)}$ & $p/p^{(0)}$ & $\chi/\chi^{(0)}$ & $\sigma_Q/\sigma_Q^{(0)}$ & $\phi_h$ & $s/\mu_Q^2$ & $\rho/\mu_Q^2$ & $p/\mu_Q^3$ & $\phi_h$ \\
\midrule
0.3 & 3.1322 & 10.035 & 1.8225 & 0.0875 & 0.3020 & 17.00 & 11.06 & 3.745 & 0.710 \\
0.6 & 3.5335 & 12.285 & 1.9993 & 0.0631 & 0.5551 & 33.72 & 22.53 & 7.622 & 0.852 \\
1.0 & 3.7688 & 13.864 & 2.1474 & 0.0488 & 0.9565 & 70.39 & 48.34 & 16.35 & 1.092 \\
1.5 & 3.8773 & 14.725 & 2.2386 & 0.0416 & 1.5348 & 158.6 & 111.6 & 37.71 & 1.501 \\
\bottomrule
\end{tabular}
\end{table}

\subsection{Transport at finite density}
\label{sec:finite:transport}

Because the equilibrium thermodynamics at fixed $\alpha$ is conformal and the gravitational sector is two-derivative Einstein gravity, the values $\eta/s=1/(4\pi)$, $c_s^2=1/2$ and $\zeta=0$ of Section~\ref{sec:transport} hold at all densities, with the same equilibrium caveat.
The incoherent conductivity follows from the horizon formula~\eqref{eq:sigmaQ} with the full values of $\phi_h$, $s$, $T$, $\mu_Q$ and $\rho$.
It decreases strongly with density (Figure~\ref{fig:finite}b and Tables~\ref{tab:finite}--\ref{tab:finite2}): at low temperature the kinematic factor gives $\sigma_Q\simeq Z(\phi_h)\,(s/\rho)^2\,(T/\mu_Q)^2/(4\pi G_4)$, accurate to $2$--$3\%$ for $T/\mu_Q\lesssim0.01$.
Because $\phi_h$ and $s/\rho$ drift slowly, the effective exponent $d\ln\sigma_Q/d\ln(T/\mu_Q)$ at the lowest temperatures reached is $1.89$--$1.95$ rather than exactly $2$.
The extremal throat accounts for this. Since $Z(\phi_h)=(48\tau_h/49)^{3/4}$ by~\eqref{eq:throat}, it predicts $\sigma_Q\propto T^2\ln^{3/4}(\mu_Q/T)$, with effective exponent $2-3/(4\tau_h+3)+\order{\tau_h^{-2}}$. For $\alpha=0.6$ at $T/\mu_Q=3\times10^{-4}$ ($\tau_h\approx6.4$) this gives $1.89$. At fixed $\mu_Q/T$ the suppression is stronger for larger $\alpha$.
The thermopower $\Theta=s/\rho$ and the thermal conductivity~\eqref{eq:kappa_bar} follow in the same way from the tabulated data.

\subsection{Verification}
\label{sec:finite:checks}

We verified the finite-density results in several independent ways; Table~\ref{tab:finite_checks} summarizes them.
At $Q=0$ the solutions reproduce the closed-form neutral values of $s/T^2$, $\sE_{\rm D}$ and the coupling $k(\alpha)$.
The Noether identity~\eqref{eq:full_noether} holds when $A_3$ is fitted directly from the asymptotic tail, and the $xx$ Einstein equation, which is not used in the integration, is satisfied.
The first law at fixed $\alpha$ holds in integrated form, $\Pi(x)-\Pi(0)=\int_0^x(\rho/T^2)\,dx'$, with $\Pi=(s/T^2+x\rho/T^2)/3$ built from the horizon entropy and the Gauss-law charge, and locally, $d\Pi/dx=\rho/T^2$, including at $\mu_Q/T\approx146$ (finite differences whose error scales as the square of the step).
Independently of the boundary condition, the Dirichlet first law $d\sE_{\rm D}=T\,ds+\mu_Q\,d\rho+\langle\mathcal O_\phi\rangle\,d\varphi_{(0)}$, with $\langle\mathcal O_\phi\rangle$ from the Ward identity, holds at finite charge to $10^{-6}$, which tests~\eqref{eq:full_EP} and the Ward identity directly.
As $\mu_Q/T\to0$ the susceptibility $\chi/T$ agrees with the closed form $16\pi e^{-\alpha/2}/(3j(\alpha))$ that follows from~\eqref{eq:chi}, and the $\order{\eps^2}$ coefficients of the horizon scalar and of $\sigma_Q$ at fixed $(S,\alpha)$ agree with the perturbative values of Tables~\ref{tab:thermo} and~\ref{tab:transport_full}, which were obtained in a different gauge and with the boundary condition imposed through the Dirichlet first law rather than directly through the Ward identity~\eqref{eq:full_k}.

\begin{table}[ht]
\centering
\caption{Verification of the nonlinear solutions: largest relative deviation over each data set, for $\alpha=0.3,0.6,1.0,1.5$. Relative deviations are independent of units.}
\label{tab:finite_checks}
\small
\renewcommand{\arraystretch}{1.15}
\begin{tabular}{lcccc}
\toprule
Check & $\alpha=0.3$ & $\alpha=0.6$ & $\alpha=1.0$ & $\alpha=1.5$ \\
\midrule
(i) neutral $s/T^2$ and $k(\alpha)$ & $3\times10^{-9}$ & $1\times10^{-9}$ & $5\times10^{-10}$ & $6\times10^{-10}$ \\
(ii) Noether identity, fitted $A_3$ & $6\times10^{-7}$ & $4\times10^{-7}$ & $2\times10^{-8}$ & $2\times10^{-8}$ \\
(iii) $xx$ equation (numerical) & \multicolumn{4}{c}{$\lesssim10^{-6}$ (satisfied identically; checked symbolically)} \\
(iv) first law (integrated) & $8\times10^{-8}$ & $2\times10^{-7}$ & $2\times10^{-7}$ & $1\times10^{-7}$ \\
(v) $\chi/T$ at $\mu_Q\to0$ from $\rho$ & $1.0\times10^{-8}$ & $1.1\times10^{-8}$ & $1.2\times10^{-8}$ & $1.2\times10^{-8}$ \\
(v) $\chi/T$ at $\mu_Q\to0$ from $s$ & $1.1\times10^{-5}$ & $3\times10^{-6}$ & $1.3\times10^{-6}$ & $3\times10^{-6}$ \\
(vi) $\delta\phi(r_h)|_{S,\alpha}$ vs.\ perturbative & $1\times10^{-5}$ & $2\times10^{-6}$ & $<10^{-5}$ & $6\times10^{-6}$ \\
(vi) $\delta\sigma_Q|_{S,\alpha}$ vs.\ perturbative & $3\times10^{-6}$ & $<10^{-6}$ & $<10^{-6}$ & $<10^{-6}$ \\
\bottomrule
\end{tabular}
\end{table}

\paragraph{Analyticity in the charge.}
The charged solutions are analytic in $Q^2$ at fixed $(\phi_h,\alpha)$. The charge enters~\eqref{eq:full_F}--\eqref{eq:full_phi} and~\eqref{eq:full_bc} only through $Q^2$, and the solution that is regular at the horizon depends analytically on $(\phi_h,Q^2)$.
The large-$r$ expansion~\eqref{eq:full_asym} contains no logarithms, because the Maxwell field sources only powers $\tilde r^{-7n-3j}$ with $n\ge1$, which never coincide with the homogeneous modes $\tilde r^{0}$ and $\tilde r^{-3}$.
The asymptotic data $\src$, $A_3$, $H_3$ and $N$ are therefore analytic in $(\phi_h,Q^2)$.
Since $\partial k/\partial\phi_h\neq0$ on the neutral family, where $dk/d\alpha=-3l\sinh\alpha/(16\pi G_4)$ and $\phi_h=\sqrt7\alpha/2$, the implicit function theorem shows that the condition~\eqref{eq:full_k} determines $\phi_h$ as an analytic function of $Q^2$.
Then $T$, $s$, $\mu_Q/Q$ and $\rho/Q$ are analytic in $Q^2$, and the pressure is analytic in $(\mu_Q/T)^2$ near zero density. In particular, non-analytic terms such as $\eps^4\ln\eps$ do not arise at any order.
The nonlinear solutions confirm this. At $\alpha=0.6$ we fitted $g\equiv\bigl(\rho/(\chi^{(0)}\mu_Q)-1\bigr)(T/\mu_Q)^2$ for $0.036\le\mu_Q/T\le0.48$ by a polynomial in $(\mu_Q/T)^2$, which reproduces $c=-0.05293$ of Table~\ref{tab:thermo} to within $10^{-5}$.
An $\eps^4\ln\eps$ term in the pressure would add a term $D\ln(\mu_Q/T)$ to $g$. Including such a term in the fit gives $|D|<3\times10^{-5}$, consistent with zero at the accuracy of the data.

\section{Charged branes in the theory with fixed source}
\label{sec:dir}

The scale-invariant boundary condition~\eqref{eq:mixedBC} used so far holds the coupling $k$ fixed. Reference~\cite{ref:original} also analyzed the neutral branes in the theory with fixed source, i.e.\ with a Dirichlet condition on $\varphi_{(0)}$, and regards that theory as the dual of the neutral solutions, because the source mode is not normalizable.
In this section we determine the charged thermodynamics, the phase structure, the transport coefficients and the hydrodynamic quasinormal modes of that theory.
The bulk solutions are the same as in the preceding sections; only their labeling changes. A charged state is labeled at fixed $\src$ rather than at fixed $\alpha$, and no boundary condition has to be solved for: every regular solution $(\phi_h,Q)$ of Section~\ref{sec:finite} is a state of the theory with fixed source.
All thermodynamic statements below rest on the Dirichlet first law $d\sE_{\rm D}=T\,ds+\mu_Q\,d\rho+\langle\mathcal O_\phi\rangle\,d\varphi_{(0)}$, which holds at finite charge (Section~\ref{sec:finite:checks}), and on the trace Ward identity~\eqref{eq:condensate}.
Unless stated otherwise we set $l=1$ and quote numbers in units $16\pi G_4=1$.

\subsection{Scale and neutral branes}
\label{sec:dir:setup}

The source introduces the energy scale
\begin{equation}
\Lambda\equiv\frac{\src^{-1/3}}{l^2},
\label{eq:Lambda}
\end{equation}
which transforms as $\Lambda\to\lambda\Lambda$ under the scaling of Section~\ref{sec:thermo:source}. The ratios $T/\Lambda$ and $\mu_Q/\Lambda$ are therefore invariant, and the thermodynamics is a function of two variables. Up to the change of frame described in Ref.~\cite{ref:original}, under which these ratios are invariant, $\Lambda$ is the scale used there. The crossover radius of Section~\ref{sec:setup} is $r\sim\src^{-1/3}=l^2\Lambda$.
Combining the Ward identity with the Smarr-type identity~\eqref{eq:full_noether} gives
\begin{equation}
3p_{\rm D}=Ts+\mu_Q\rho+\varphi_{(1)}\langle\mathcal O_\phi\rangle,\qquad
dp_{\rm D}=s\,dT+\rho\,d\mu_Q-\langle\mathcal O_\phi\rangle\,d\varphi_{(0)},
\label{eq:dir_ward}
\end{equation}
so that $p_{\rm D}=\Lambda^3\,\Pi_{\rm D}(T/\Lambda,\mu_Q/\Lambda)$ with a single dimensionless function $\Pi_{\rm D}$.

For the neutral branes $\src=\sinh\alpha/(ml^2)$ and $r_h=(2ml^2)^{1/3}$ give
\begin{equation}
\frac{T}{\Lambda}=\frac{3}{4\pi}\,(2\sinh\alpha)^{1/3}\,e^{-\alpha/2},
\label{eq:dir_TL}
\end{equation}
which has a maximum $T_{\max}\simeq0.193803\,\Lambda$ at $\alpha_*=\tfrac12\ln5$ and vanishes as $\alpha\to0$ and $\alpha\to\infty$~\cite{ref:original}.
Each $T<T_{\max}$ is thus reached by two branes. On the small-$\alpha$ branch the specific heat is positive; on the large-$\alpha$ branch it is negative and $c_s^2=(2\coth\alpha-3)/(4\coth\alpha+3)<0$~\cite{ref:original}.
By~\eqref{eq:eps_p}, $p_{\rm D}>0$ for $\tanh(\alpha/2)<2/3$, i.e.\ for $\alpha<\ln5$, which includes the whole small-$\alpha$ branch.

\subsection{Susceptibility and second-order thermodynamics}
\label{sec:dir:second}

\paragraph{First order.}
The susceptibility~\eqref{eq:chi} does not depend on the boundary condition. Combined with~\eqref{eq:neutral_thermo} and~\eqref{eq:dir_TL} it gives $\chi(T/\Lambda)$ in parametric form,
\begin{equation}
\frac{\chi}{T}=\frac{e^{-\alpha/2}}{3G_4\,j(\alpha)} .
\label{eq:dir_chi}
\end{equation}
At low temperature on the stable branch, \eqref{eq:j_small} gives
\begin{equation}
\frac1\chi=\frac{3G_4}{T}-\frac{8\pi^2G_4}{3\sqrt3\,\Lambda}+\order{\frac{T^2}{\Lambda^3}} .
\label{eq:dir_chi_lowT}
\end{equation}
The correction to the infrared value $\chi=T/(3G_4)$ is therefore linear in $T/\Lambda$, whereas the neutral thermodynamics is corrected at order $(T/\Lambda)^3$~\cite{ref:original}. It is a $T$-independent decrease of $1/\chi$, the electrostatic potential drop per unit charge: in the region $r\gtrsim\src^{-1/3}$ the coupling $Z(\phi_0)$ is large and the potential no longer drops.
Near $T_{\max}$ the two branches meet with a square-root singularity,
$16\pi G_4\,\chi/T\simeq204.16\,\bigl[1\mp3.732\,(1-T/T_{\max})^{1/2}\bigr]$, with the upper sign on the stable branch, so that $d\chi/dT$ diverges there.
On the unstable branch $16\pi G_4\,\chi/T\to(64\pi/3)\,(3\Lambda/4\pi T)^9$ as $T\to0$.
The diffusion constant obeys~\eqref{eq:DQ} unchanged, with $\alpha$ now a function of $T/\Lambda$: $4\pi TD_Q/3$ decreases from $1$ at $T\to0$ to $0.2744$ at $T_{\max}$ and tends to $\tfrac14$ at the low-temperature end of the unstable branch.

\paragraph{Second order.}
Without a boundary condition to impose, the first law gives directly
\begin{equation}
\sE_{\rm D}(S,\rho;\varphi_{(0)})=\sE_{\rm D}^{(0)}(S;\varphi_{(0)})+\frac{\rho^2}{2\chi(S;\varphi_{(0)})}+\order{\rho^4},
\label{eq:dir_E}
\end{equation}
from which $T^{(2)}|_{S,\src}=(8\pi G_4l)^{-1}\partial\mu_Q^{(1)}/\partial S|_{\src}$ (Section~\ref{sec:thermo:check}) and $\langle\mathcal O_\phi\rangle^{(2)}|_{S,\src}=\partial_{\varphi_{(0)}}(\rho^2/2\chi)|_S$ follow.
At fixed temperature the pressure is $p_{\rm D}=p_{\rm D}^{(0)}(T)+\tfrac12\chi(T)\,\mu_Q^2+\order{\mu_Q^4}$, and with $L\equiv d\ln\chi/d\ln T|_{\src}$,
\begin{equation}
\frac{\Delta s}{s^{(0)}}=\frac{L\,\chi T}{2s^{(0)}}\,\frac{\mu_Q^2}{T^2},\qquad
\Delta\langle\mathcal O_\phi\rangle=\frac{(1-L)\,\chi\,\mu_Q^2}{2\varphi_{(1)}},\qquad
L=\frac{\tfrac13\coth\alpha-1-j'(\alpha)/j(\alpha)}{\tfrac13\coth\alpha-\tfrac12}.
\label{eq:dir_fixedT}
\end{equation}
If $\chi$ were proportional to $T$, as at the infrared fixed point, $L$ would equal $1$ and the condensate would not shift at this order. Here $L$ grows from $1$ at $T\to0$ and diverges at $T_{\max}$, where $\partial T/\partial\alpha|_{\src}=0$; on the unstable branch it is negative and tends to $-8$.
The horizon scalar and $\sigma_Q$ require the second-order solution. They follow from the data of Table~\ref{tab:thermo} by a relabeling at fixed entropy: the source shift $\delta\varphi_{(0)}$ is removed by $\delta\alpha=-\delta\varphi_{(0)}/[\tfrac{\sqrt7}{2}(\coth\alpha+\tfrac34)]$, so that $\delta\phi(r_h)|_{S,\src}=\delta\phi(r_h)|_{S,\alpha}+\tfrac{\sqrt7}{2}\delta\alpha$ and $\delta\sigma_Q/\sigma_Q^{(0)}|_{S,\src}=\delta\sigma_Q/\sigma_Q^{(0)}|_{S,\alpha}+\tfrac32\delta\alpha$.
At fixed source the horizon scalar is shifted upward, $\delta\phi(r_h)|_{S,\src}=+0.00456$ at $\alpha=0.6$ (Table~\ref{tab:conv}), opposite to its shift at fixed $\alpha$.
Table~\ref{tab:dir_second} lists the resulting coefficients at fixed $(T,\Lambda)$.

\begin{table}[ht]
\centering
\caption{Theory with fixed source: susceptibility, diffusion constant and coefficients of $(\mu_Q/T)^2$ at fixed $(T,\Lambda)$, in units $16\pi G_4=l=1$. The last column is the quartic coefficient~\eqref{eq:dir_c}. The first six rows lie on the stable branch ($\alpha<\alpha_*\simeq0.805$), the last three on the unstable branch.}
\label{tab:dir_second}
\small
\setlength{\tabcolsep}{4.5pt}
\renewcommand{\arraystretch}{1.15}
\begin{tabular}{cccccccc}
\toprule
$\alpha$ & $T/\Lambda$ & $16\pi G_4\,\chi/T$ & $4\pi TD_Q/3$ & $L$ & $\Delta s/s^{(0)}$ & $\Delta\sigma_Q/\sigma_Q^{(0)}$ & $c_{\rm D}$ \\
\midrule
0.05 & 0.1081 & 34.06 & 0.5302 & 1.925 & 0.1379 & $-0.0963$ & 0.1219 \\
0.1  & 0.1329 & 43.01 & 0.4526 & 2.388 & 0.2005 & $-0.0984$ & 0.1929 \\
0.2  & 0.1595 & 59.67 & 0.3790 & 3.366 & 0.3374 & $-0.0806$ & 0.3509 \\
0.4  & 0.1831 & 96.22 & 0.3173 & 6.579 & 0.7878 & $\phantom{-}0.0175$ & 0.8687 \\
0.6  & 0.1917 & 142.3 & 0.2897 & 16.24 & 2.130  & $\phantom{-}0.2889$ & 2.398 \\
0.8  & 0.1938 & 202.5 & 0.2747 & 862.4 & 119.3  & $\phantom{-}21.12$  & 135.4 \\
\midrule
1.0  & 0.1925 & 282.5 & 0.2658 & $-25.15$ & $-3.595$ & $-0.655$ & $-4.097$ \\
1.5  & 0.1828 & 622.2 & 0.2555 & $-10.72$ & $-1.595$ & $-0.224$ & $-1.822$ \\
2.0  & 0.1700 & 1335.5 & 0.2520 & $-8.829$ & $-1.331$ & $-0.122$ & $-1.521$ \\
\bottomrule
\end{tabular}
\end{table}

\subsection{Quartic response and the maximal temperature}
\label{sec:dir:quartic}

Write $\rho=\chi\,\mu_Q\bigl[1+c_{\rm D}\,\mu_Q^2/T^2+\order{\mu_Q^4}\bigr]$ at fixed $(T,\Lambda)$, so that $T^2\chi_4/\chi_2=6c_{\rm D}$ as in~\eqref{eq:chi4}. Expanding the solution with unshifted source at fixed $(S,\src)$ gives
\begin{equation}
c_{\rm D}=-\Bigl(\frac{T^{(0)}}{\mu_Q^{(1)}}\Bigr)^2\Bigl[L\,\frac{T^{(2)}}{T^{(0)}}\bigg|_{S,\src}+\frac{\mu_Q^{(3)}}{\mu_Q^{(1)}}\bigg|_{S,\src}\Bigr]
\equiv c_{\rm D}^{\rm th}+c_{\rm D}^{\rm s},
\label{eq:dir_c}
\end{equation}
with $\mu_Q^{(3)}|_{S,\src}$ given by~\eqref{eq:mu3}.
The first term is known in closed form. The second measures the departure of $\mu_Q(S,\rho)$ from linearity at fixed source, which vanishes for the Reissner--Nordstr\"om brane and here comes from the backreaction of the scalar.
It is small: $c_{\rm D}^{\rm s}$ is positive, has a maximum $0.0036$ near $\alpha\simeq0.05$--$0.1$, and decreases to $2\times10^{-5}$ at $\alpha=1.5$.
As $T\to0$ it tends to $0.00125$, a value we obtained independently from the linear scalar sourced by $Z'(0)F^2$ on $\AdS_4$-Schwarzschild, in which the scalar is dual to the operator with $\Delta=6$.
Unlike $c<0$ at fixed $\alpha$, $c_{\rm D}$ is positive on the stable branch and negative on the unstable one (Table~\ref{tab:dir_second}), with the limits
\begin{equation}
c_{\rm D}\simeq\frac{3}{16\pi^2}\Bigl(1-\frac{2\pi}{3\sqrt3}\,\frac{4\pi T}{3\Lambda}\Bigr)^{-3}+0.00125\quad(T\to0),\qquad
c_{\rm D}\to-\frac{96}{7\pi^2}\quad(\text{unstable branch},\ T\to0).
\label{eq:dir_c_limits}
\end{equation}
In the first limit the closed-form part tends to the Reissner--Nordstr\"om value $3/(16\pi^2)$ with corrections of relative order $(T/\Lambda)^3$ beyond those shown, while the order-$T/\Lambda$ correction to $c_{\rm D}^{\rm s}$ is not known; $c_{\rm D}^{\rm s}$ is $0.0025$ at $T/\Lambda=0.05$.

At $T_{\max}$ the coefficient diverges, $c_{\rm D}\simeq\pm0.2934\,(1-T/T_{\max})^{-1/2}$, positive on the stable branch.
The divergence comes from $L$ in~\eqref{eq:dir_c} and is the expansion of a shift of the maximal temperature. Since $\partial T/\partial S|_{\src}=0$ at $\alpha_*$, the shift of the maximum at fixed $\mu_Q$ follows from the closed-form $T^{(2)}|_{S,\src}$,
\begin{equation}
T_{\max}(\mu_Q)=T_{\max}\Bigl[1-\kappa_T\,\frac{\mu_Q^2}{T_{\max}^2}+\order{\mu_Q^4}\Bigr],\qquad
\kappa_T=-\Bigl(\frac{T^{(0)}}{\mu_Q^{(1)}}\Bigr)^2\frac{T^{(2)}}{T^{(0)}}\bigg|_{S,\src,\alpha_*}=0.15723 .
\label{eq:dir_Tmax}
\end{equation}
Near $T_{\max}$ the susceptibility on the stable branch is
\begin{equation}
\chi\simeq\chi_*\Bigl[1-g\,\sqrt{2\bigl(1-T/T_{\max}(\mu_Q)\bigr)/\kappa_2}\Bigr],\qquad g=\frac{d\ln\chi}{d\alpha}\Big|_{\alpha_*},\qquad \kappa_2=-\frac{d^2\ln T}{d\alpha^2}\Big|_{\alpha_*}=\frac{5}{12},
\end{equation}
and expanding in $\mu_Q^2$ at fixed $T$ reproduces the coefficient $0.2934=g\,\kappa_T/\sqrt{2\kappa_2}$.
The expansion in $\mu_Q/T$ therefore has a radius of convergence that shrinks to zero at $T_{\max}$, set by $\mu_Q^2\lesssim(1-T/T_{\max})\,T_{\max}^2/\kappa_T$.

We verified these results in three ways. The two routes to $c_{\rm D}$, the relabeling of Table~\ref{tab:thermo} and a Legendre transform of the fixed-$k$ energy, agree to ten digits.
Nonlinear solutions held at fixed $(S,\src)$ reproduce $T^{(2)}|_{S,\src}$ to seven digits for $\alpha=0.3$--$2$, $\delta\phi(r_h)|_{S,\src}$ to three digits and $c_{\rm D}^{\rm s}$ to $5\times10^{-6}$.
Nonlinear solutions at fixed $T/\Lambda$ give $c_{\rm D}=0.8687$, $2.3978$ and $-4.0959$ at $\alpha=0.4$, $0.6$ and $1.0$, compared with $0.8687$, $2.3985$ and $-4.0969$ from~\eqref{eq:dir_c}; and maximizing $T/\Lambda$ at fixed $\mu_Q/\Lambda$ gives $\kappa_T=0.15721$ at $\mu_Q/T_{\max}=0.10$ and $0.15716$ at $0.21$.

\subsection{Phase structure at finite density}
\label{sec:dir:phase}

\begin{figure}[ht]
\centering
\includegraphics[width=0.98\textwidth]{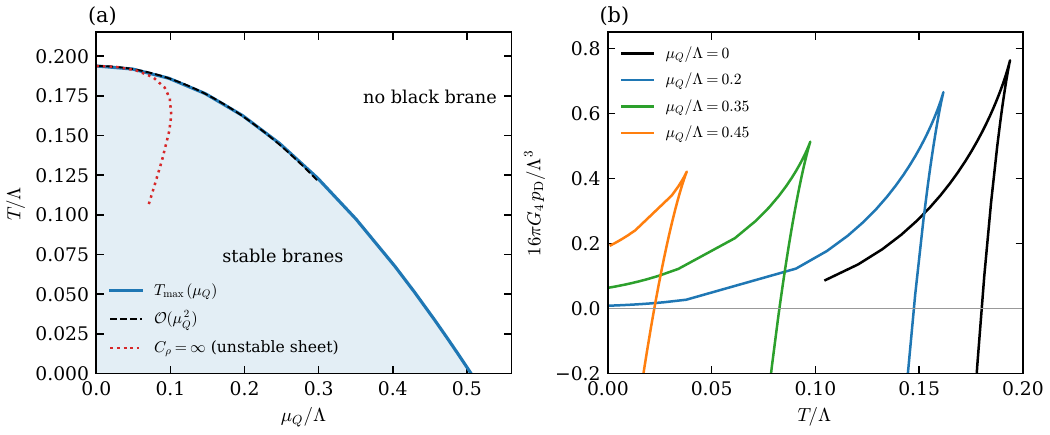}
\caption{Theory with fixed source. (a)~Region of existence of charged black branes. The boundary is the maximal temperature $T_{\max}(\mu_Q)$, which ends at $\mu_{\max}=0.50565\,\Lambda$; the dashed curve is the quadratic approximation~\eqref{eq:dir_Tmax}, and the dotted curve, which lies on the unstable branch, is where $C_\rho$ diverges. (b)~Pressure $p_{\rm D}=-\mathfrak f_{\rm D}$ of both branches at fixed $\mu_Q/\Lambda$; the two branches meet at a cusp at $T_{\max}(\mu_Q)$, and the stable branch has the higher pressure. Units $16\pi G_4=l=1$.}
\label{fig:dir_phase}
\end{figure}

To go beyond small $\mu_Q$ we mapped the regular solutions $(\phi_h,Q)$, at $r_h=1$, onto the plane $(\mu_Q/\Lambda,T/\Lambda)$.
The image is a bounded region (Figure~\ref{fig:dir_phase}a): black branes with fixed source exist only for $T<T_{\max}(\mu_Q)$, and every point inside is reached by exactly two solutions, which merge on the boundary.
The boundary curve is the fold of the map from $(s,\rho)$ to $(T,\mu_Q)$, on which the Jacobian $\det\partial(T,\mu_Q)/\partial(s,\rho)$ vanishes. We located it by solving for the zero of this determinant along lines of fixed $Q/Q_{\max}(\phi_h)$ and checked it by maximizing $T/\Lambda$ at fixed $\mu_Q/\Lambda$, which agrees to six digits at $\mu_Q/\Lambda=0.15$ and $0.30$.
Selected points are
\begin{center}
\small
\setlength{\tabcolsep}{4pt}
\begin{tabular}{c|ccccccccccc}
$\mu_Q/\Lambda$ & 0 & 0.098 & 0.197 & 0.299 & 0.350 & 0.403 & 0.456 & 0.483 & 0.499 & 0.505 & 0.5057 \\
\hline
$T_{\max}/\Lambda$ & 0.1938 & 0.1861 & 0.1626 & 0.1232 & 0.0974 & 0.0676 & 0.0339 & 0.0158 & 0.0051 & 0.0005 & 0
\end{tabular}
\end{center}
The quadratic form~\eqref{eq:dir_Tmax} is accurate up to $\mu_Q\approx0.3\Lambda$. Near its end the curve is linear, $T_{\max}\simeq0.71\,(\mu_{\max}-\mu_Q)$, and it reaches $T=0$ at
\begin{equation}
\mu_{\max}=0.50565\,\Lambda ,
\label{eq:dir_mumax}
\end{equation}
the largest chemical potential of any zero-temperature solution (Section~\ref{sec:dir:T0}). Along the curve $\rho/s$ increases from $0$ to the extremal value $\sqrt{21}/(2\pi)$.

\paragraph{Local stability.}
Thermodynamic stability of a translationally invariant brane requires the Hessian of $\sE_{\rm D}(s,\rho)$ at fixed source, i.e.\ $M\equiv\partial(T,\mu_Q)/\partial(s,\rho)|_{\src}$, to be positive definite.
We find that $M$ is positive definite throughout the region bounded by $T_{\max}(\mu_Q)$ on the branch continuously connected to the small-$\alpha$ branes, and has one negative eigenvalue on the other branch; $\det M$ changes sign only on the fold.
On the stable branch the pressure is positive and larger than on the unstable branch at the same $(T,\mu_Q)$ (Figure~\ref{fig:dir_phase}b), and we found no further folds or crossings, so there is no first-order transition between black branes. As at zero density, no black brane with the given source exists above $T_{\max}(\mu_Q)$.
At $\mu_Q\neq0$ the speed of sound, $c_s^2=v^{\rm T}Mv/(\sE+p)$ with $v=(s,\rho)$, is positive on the fold, with a maximum of $0.0083$ at $\mu_Q\approx0.35\Lambda$; it vanishes on a nearby curve inside the unstable branch, and the two curves meet only at $\mu_Q=0$.
Since $\chi=(M^{-1})_{22}=M_{11}/\det M$, the susceptibility diverges on the fold at $\mu_Q\neq0$, whereas at $\mu_Q=0$, where $M_{11}\to0$ instead, only $d\chi/dT$ diverges. What vanishes on the fold is the diffusion constant of the coupled heat and charge mode. Linearized hydrodynamics with the conserved current $J^i=\rho v^i-\sigma_QT\partial^i(\mu_Q/T)$ gives
\begin{equation}
\omega=-iDk^2,\qquad D=\frac{\sigma_Q\,(\sE+p)\,\det M}{T^2\,c_s^2},
\label{eq:dir_D}
\end{equation}
which reduces to $D_Q=\sigma_Q/\chi$ at $\mu_Q=0$. At finite density the instability at the fold therefore sets in through this diffusive mode, while sound is still stable; at $\mu_Q=0$ both set in together. Section~\ref{sec:dir:qnm} confirms this with quasinormal modes.
The specific heat at fixed charge, $C_\rho=T/M_{11}$, diverges on a different curve, which lies on the unstable branch (dotted in Figure~\ref{fig:dir_phase}a); in the canonical ensemble part of the unstable branch therefore has $C_\rho>0$, but these states are unstable under the full Hessian.

\subsection{Hydrodynamics and transport}
\label{sec:dir:hydro}

The theory with fixed source is not conformal, and its first-order hydrodynamics involves $c_s^2$, $\zeta$, $\sigma_Q$ and the diffusion constant~\eqref{eq:dir_D}, all of which depend on $(T/\Lambda,\mu_Q/\Lambda)$; $\eta/s=1/(4\pi)$ at all densities.
We evaluate $c_s^2$ and $D$ from the Hessian $M$, $\sigma_Q$ from~\eqref{eq:sigmaQ}, and $\zeta$ from the Eling--Oz formula~\cite{ref:EO}, $\zeta/\eta=\bigl(s\,\partial_s\phi_h+\rho\,\partial_\rho\phi_h\bigr)^2$ with derivatives at fixed source, which reproduces $\zeta/\eta=63/(4\coth\alpha+3)^2$ of Ref.~\cite{ref:original} at $\mu_Q=0$.
Table~\ref{tab:dir_hydro} gives representative values.
At fixed temperature $c_s^2$ decreases with $\mu_Q$ and becomes small near $T_{\max}(\mu_Q)$; at small density $dc_s^2/d(\mu_Q/T)^2=-0.042$, $-0.082$ and $-0.154$ at $T/\Lambda=0.10$, $0.14$ and $0.17$, from the closed-form pressure of Section~\ref{sec:dir:second}.
The ratio $\zeta/\eta$ grows with $\mu_Q$, $\sigma_Q$ falls (mostly through the kinematic factor in~\eqref{eq:sigmaQ}), and $TD$ falls and vanishes on the fold.

Buchel's conjectured bound~\cite{ref:BuchelBound}, $\zeta/\eta\ge2(1/d_s-c_s^2)$ with $d_s=2$, is violated at zero density for all $\alpha$~\cite{ref:original}; violations in other holographic models are discussed in Ref.~\cite{ref:Yarom}.
At finite density two forms have been considered: the one with the speed of sound, and one with $c_q^2\equiv(\partial p/\partial\sE)_\rho=(sM_{11}+\rho M_{12})/T$ proposed in Ref.~\cite{ref:GSSST}.
With $c_s^2$ the bound is violated at every point we computed; the ratio $R_s=(\zeta/\eta)/[2(\tfrac12-c_s^2)]$ first decreases with $\mu_Q$ and then increases, down to $0.004$.
With $c_q^2$ it is violated except at low temperature and large density, where $c_q^2>\tfrac12$ and the bound holds trivially (for instance at $T/\Lambda=0.06$ and $\mu_Q\ge0.3\Lambda$, and at all $\mu_Q\ge0.2\Lambda$ as $T\to0$).
Since the two forms agree at $\mu_Q=0$, where the bound is already violated, the density does not restore it in either form.

\begin{table}[ht]
\centering
\caption{Hydrodynamic data of the theory with fixed source on the stable branch: speed of sound, bulk viscosity, the Buchel ratios $R_s$ and $R_q$ (bound satisfied if $\ge1$; ``triv.'' where $c_q^2>\tfrac12$), incoherent conductivity relative to its value at the same $T$ and $\mu_Q=0$, and the diffusion constant~\eqref{eq:dir_D}.}
\label{tab:dir_hydro}
\small
\setlength{\tabcolsep}{5pt}
\renewcommand{\arraystretch}{1.15}
\begin{tabular}{cccccccc}
\toprule
$T/\Lambda$ & $\mu_Q/\Lambda$ & $c_s^2$ & $\zeta/\eta$ & $R_s$ & $R_q$ & $\sigma_Q/\sigma_Q|_{\mu_Q=0}$ & $TD$ \\
\midrule
0.06 & 0    & 0.4910 & 0.00025 & 0.014 & 0.014 & 1      & 0.1703 \\
0.06 & 0.10 & 0.4479 & 0.00040 & 0.0038 & 0.023 & 0.628 & 0.1439 \\
0.06 & 0.20 & 0.3657 & 0.00125 & 0.0046 & 0.107 & 0.276 & 0.1007 \\
0.06 & 0.30 & 0.2654 & 0.00605 & 0.0129 & triv. & 0.137 & 0.0656 \\
0.06 & 0.40 & 0.0995 & 0.0789  & 0.099 & triv. & 0.087 & 0.0346 \\
0.10 & 0    & 0.4574 & 0.00563 & 0.066 & 0.066 & 1      & 0.1332 \\
0.10 & 0.20 & 0.3262 & 0.0174  & 0.050 & 0.160 & 0.486 & 0.0942 \\
0.10 & 0.30 & 0.1830 & 0.0788  & 0.124 & 0.411 & 0.302 & 0.0612 \\
0.14 & 0    & 0.3762 & 0.0477  & 0.193 & 0.193 & 1      & 0.1031 \\
0.14 & 0.20 & 0.2136 & 0.1437  & 0.251 & 0.375 & 0.669 & 0.0738 \\
0.17 & 0    & 0.2554 & 0.1861  & 0.381 & 0.381 & 1      & 0.0840 \\
0.17 & 0.15 & 0.1134 & 0.3992  & 0.516 & 0.574 & 0.978 & 0.0665 \\
\bottomrule
\end{tabular}
\end{table}

\begin{figure}[ht]
\centering
\includegraphics[width=0.98\textwidth]{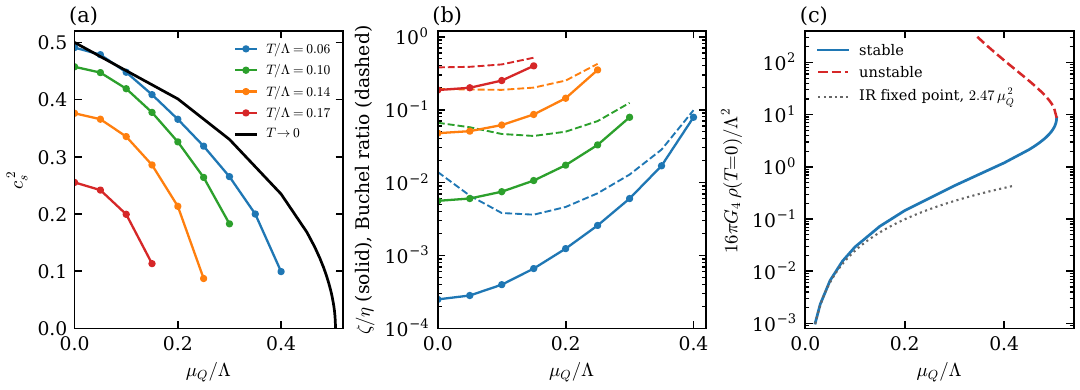}
\caption{Theory with fixed source on the stable branch. (a)~Speed of sound at fixed $T/\Lambda$ as a function of $\mu_Q/\Lambda$, and its zero-temperature limit $c_s^2=\rho_0/(\mu_Q\,d\rho_0/d\mu_Q)$, which vanishes at $\mu_{\max}$. (b)~Bulk viscosity from the Eling--Oz formula (solid) and the Buchel ratio $R_s$ (dashed); the bound requires $R_s\ge1$. (c)~Zero-temperature charge density of the stable and unstable branches; the dotted line is the infrared value $a\mu_Q^2$, and the stable branch follows~\eqref{eq:dir_lowmu}. Units $16\pi G_4=l=1$.}
\label{fig:dir_hydro}
\end{figure}

\subsection{Zero temperature}
\label{sec:dir:T0}

The infrared throat of Proposition~\ref{prop:throat} does not depend on the boundary condition, and both branches end in it as $T\to0$ at fixed $\mu_Q<\mu_{\max}$.
Along paths of fixed $\mu_Q/\Lambda$ between $0.2$ and $0.45$ the local exponent~\eqref{eq:local_exp} approaches its prediction as at fixed $\alpha$, with a ratio of measured to predicted values that grows from $0.78$ to $0.92$ as $\tau_h$ increases, and $\rho/s$ approaches the horizon value of the bound from below.
What is new is the zero-temperature equation of state $p_0(\mu_Q)=\lim_{T\to0}p_{\rm D}$, a function of $\mu_Q/\Lambda$, and $\rho_0=dp_0/d\mu_Q$.
We obtained it from solutions down to $T/\mu_Q\sim10^{-8}$, using $p_0\simeq p_{\rm D}-Ts$, whose error is of relative order $Ts/(p\,\tau_h^2)$:
\begin{center}
\small
\setlength{\tabcolsep}{4pt}
\begin{tabular}{c|ccccccccc}
$\mu_Q/\Lambda$ & 0.05 & 0.1 & 0.2 & 0.3 & 0.4 & 0.45 & 0.48 & 0.50 & 0.505 \\
\hline
$16\pi G_4\,p_0/\Lambda^3$ & 0.000109 & 0.000932 & 0.00868 & 0.0356 & 0.1104 & 0.1904 & 0.2706 & 0.3556 & 0.3871 \\
$16\pi G_4\,\rho_0/\Lambda^2$ & 0.006685 & 0.02923 & 0.1449 & 0.4351 & 1.193 & 2.134 & 3.377 & 5.561 & $\approx7.5$ \\
$c_s^2(T\!=\!0)$ & -- & -- & 0.401 & 0.331 & 0.235 & 0.168 & 0.11 & 0.05 & 0.02
\end{tabular}
\end{center}
The speed of sound at $T=0$ is $c_s^2=\rho_0/(\mu_Q\,d\rho_0/d\mu_Q)$; it decreases from $\tfrac12$ to $0$ at $\mu_{\max}$, where $d\rho_0/d\mu_Q$ diverges and $\rho_0(\mu_Q)$ joins the unstable branch with a square-root singularity. On the unstable branch $\rho_0$ is much larger ($13.8$ at $\mu_Q=0.50\Lambda$, $110$ at $0.40\Lambda$ and $1.4\times10^4$ at $0.20\Lambda$) and $p_0$ is negative below $\mu_Q\simeq0.48\Lambda$. As $T\to0$ at fixed $\mu_Q$ on the stable branch, $\zeta/\eta$ vanishes (roughly as $T^{1.5}$ in the range computed), $\sigma_Q\propto T^2$ up to the logarithm of Section~\ref{sec:finite:transport}, and $D\to0$.

\paragraph{Low density.}
For $\mu_Q\ll\Lambda$ the brane lies deep in the infrared $\AdS_4$ region, and the only effect of the ultraviolet at leading order is the shift of $1/\chi$ in~\eqref{eq:dir_chi_lowT}, which applies to the electrostatic potential drop of any state with $r_h\ll\src^{-1/3}$. This gives
\begin{equation}
\rho_0=a\,\Bigl(\mu_Q+\frac{8\pi^2G_4}{3\sqrt3}\,\frac{\rho_0}{\Lambda}\Bigr)^2,\qquad 16\pi G_4\,a=2.4708,
\label{eq:dir_lowmu}
\end{equation}
where $a$ is the zero-temperature value of $\rho/\mu_Q^2$ of the charged branes of the infrared fixed point, i.e.\ of the source-free solutions with $\src=0$, which we computed separately; the Reissner--Nordstr\"om value would be $16\pi G_4a=4/\sqrt3$.
Equation~\eqref{eq:dir_lowmu} has no adjustable parameter. It reproduces the numerical $\rho_0$ to $8\times10^{-6}$ at $\mu_Q=0.02\Lambda$, $2\times10^{-4}$ at $0.05\Lambda$, $1.7\times10^{-3}$ at $0.1\Lambda$ and $2\%$ at $0.2\Lambda$, with deviations that scale as $(\mu_Q/\Lambda)^3$; a free fit of both coefficients at $\mu_Q\le0.075\Lambda$ gives $2.4717$ and $0.2999$ in place of $2.4708$ and $\pi/(6\sqrt3)=0.3023$.
The shift of the potential drop makes $\mu_Q(\rho_0)$ non-monotonic: \eqref{eq:dir_lowmu} alone gives a maximal chemical potential $\Lambda/(4a\cdot8\pi^2G_4/3\sqrt3)\simeq0.33\,\Lambda$, which accounts qualitatively for the existence of $\mu_{\max}$, while the full value~\eqref{eq:dir_mumax} also involves the higher-order terms.

\paragraph{Matching constant.}
The throat leaves one integration constant, the shift of $\tau$ (Appendix~\ref{app:throat}), which the ultraviolet fixes.
We define $\tau_E\equiv\tfrac{49}{48}\,e^{4\phi_h/\sqrt7}$. With the next-to-leading-order relation
$e^{-4\phi/\sqrt7}=\tfrac{49}{48\tau}\bigl(1+\tfrac{9}{16}\tau^{-1}\ln\tau+\cdots\bigr)$, which follows from~\eqref{eq:app_NLO}, the near-extremal temperature takes the form
\begin{equation}
\frac{T}{\mu_Q}=\frac{21}{4\pi}\,e^{\mathcal C}\,e^{-3\phi_h/\sqrt7}\,\tau_E^{-9/16}\,e^{-\tau_E}\bigl[1+o(1)\bigr],
\label{eq:dir_C}
\end{equation}
where $\mathcal C$ collects the shift of $\tau$ and the ratio of throat time to boundary time. In the theory with fixed $\alpha$ it is a function of $\alpha$; here it is a function of $\mu_Q/\Lambda$ and of the branch.
Evaluated on our solutions, $\mathcal C$ converges to three digits as $T/\mu_Q$ decreases from $10^{-3}$ to $10^{-8}$. On the stable branch $\mathcal C=-1.825$, $-1.755$, $-1.642$, $-1.473$, $-1.164$, $-0.846$, $-0.459$ and $+0.206$ at $\mu_Q/\Lambda=0.02$, $0.1$, $0.2$, $0.3$, $0.4$, $0.45$, $0.48$ and $0.50$.
On the unstable branch it is larger, $\mathcal C=2.93$ at $\mu_Q=0.50\Lambda$ and $7.35$ at $0.48\Lambda$, and it grows rapidly at smaller $\mu_Q$ (about $78$ at $0.40\Lambda$ and $3.1\times10^5$ at $0.20\Lambda$). There the horizon scalar stays close to a value fixed by $\mu_Q/\Lambda$ ($\phi_h\simeq2.9$ and $8.35$ at $\mu_Q/\Lambda=0.4$ and $0.2$), and the logarithmic running of the throat becomes visible only for $T/\mu_Q\lesssim e^{-\mathcal C}$.

\subsection{Quasinormal modes and the asymptotic region}
\label{sec:dir:qnm}

The hydrodynamic predictions above, like those of Section~\ref{sec:transport:modes}, assume that the long-wavelength poles of the dual theory are those of first-order hydrodynamics. We tested this by computing the quasinormal modes of the charged branes.

\paragraph{Method.}
We derived the linearized field equations in the radial gauge for the sound channel, with perturbations $h_{tt},h_{tx},h_{xx},h_{yy},\delta\phi,a_t,a_x\propto e^{-i\omega t+ikx}$, and for the transverse channel $h_{ty},h_{xy},a_y$.
At each radius we reduced them numerically to closed equations for the gauge-invariant combinations~\cite{ref:KovtunStarinets,ref:BBS}: in the sound channel $Z_H=-k^2(a^2/h^2)\hat h_{tt}+2\omega k\,\hat h_{tx}+\omega^2\hat h_{xx}+\bigl[k^2(a^2)'/(h^2)'-\omega^2\bigr]\hat h_{yy}$, $Z_\phi=\delta\phi-\phi'\,h^2\hat h_{yy}/(h^2)'$ and $Z_A=\omega a_x+ka_t-kA_t'h^2\hat h_{yy}/(h^2)'$, where $\hat h_{tt}=-h_{tt}/a^2$ and $\hat h_{ij}=h_{ij}/h^2$; and in the transverse channel $k\hat h_{ty}+\omega\hat h_{xy}$ and $a_y$. The reduction closes to relative residuals of $10^{-15}$--$10^{-11}$, which checks the linearized equations.
We impose ingoing conditions at the horizon.
At large $r$, where the metric is~\eqref{eq:UVmetric}, every channel reduces to a Bessel equation in $2\sqrt{k^2-\omega^2}\,\src^{1/2}r^{1/2}$ (with $l=1$). For spacelike momenta, $\operatorname{Re}\sqrt{k^2-\omega^2}>0$, which includes all hydrodynamic modes because $c_s^2<1$, one solution grows exponentially and is not normalizable, and the other decays; we require the decaying one. The radiating region of Appendix~\ref{app:dynamics} occurs only for $|\omega|>k$.
We locate the modes from the determinant of ingoing and decaying solutions matched at an intermediate radius.

\begin{figure}[ht]
\centering
\includegraphics[width=0.98\textwidth]{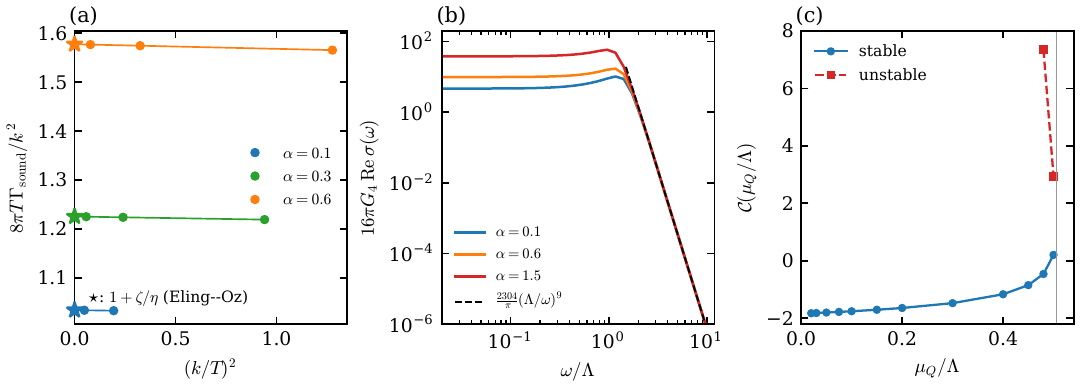}
\caption{(a)~Attenuation of the sound quasinormal mode at zero density as a function of $(k/T)^2$; the stars are $1+\zeta/\eta$ from the Eling--Oz formula. (b)~Optical conductivity at zero density with the source defined in Section~\ref{sec:dir:qnm}; the dashed line is the asymptote $(2304/\pi)(\Lambda/\omega)^9$. (c)~Matching constant $\mathcal C$ of~\eqref{eq:dir_C}; on the unstable branch it exceeds the range shown for $\mu_Q\lesssim0.47\Lambda$. Units $16\pi G_4=l=1$.}
\label{fig:dir_modes}
\end{figure}

\paragraph{Results.}
At zero density, on the exact neutral branes, the charge-diffusion and shear poles satisfy $\Gamma/(D k^2)\to1$, with $D=D_Q$ of~\eqref{eq:DQ} and $D=1/(4\pi T)$ respectively: for $\alpha=0.3$--$1.5$, including the unstable branch, the deviation is at most $6\times10^{-5}$ at the smallest wavevectors, $k/T=0.1$--$0.2$, and grows as $k^2$.
The sound pole reproduces the speed of sound and the attenuation $\Gamma_s=(\eta+\zeta)/(2sT)$ of the theory with fixed source:
\begin{center}
\small
\setlength{\tabcolsep}{5pt}
\begin{tabular}{cccccc}
$\alpha$ & $k/T$ & $\operatorname{Re}\omega/k$ & $c_s$ (thermodynamics) & $8\pi T\,\Gamma_s/k^2$ & $1+\zeta/\eta$ (Eling--Oz) \\
\hline
0.1 & 0.22 & 0.62920 & 0.62903 & 1.0336 & 1.0339 \\
0.3 & 0.24 & 0.48087 & 0.48066 & 1.2247 & 1.2251 \\
0.6 & 0.28 & 0.26360 & 0.26325 & 1.5763 & 1.5771
\end{tabular}
\end{center}
with deviations that decrease as $k^2$ (Figure~\ref{fig:dir_modes}a). On the unstable branch the sound pole moves to the positive imaginary axis, $\omega\simeq i|c_s|k-i(\eta+\zeta)k^2/(2sT)$; at $\alpha=1.0$ and $k/T=0.35$ we find $\operatorname{Im}\omega=0.009305$ against $0.009321$, and at $\alpha=0.9$ and $1.5$ the agreement is equally close. The large-$\alpha$ branes are therefore dynamically unstable, as anticipated in Ref.~\cite{ref:original} and as the correlated stability conjecture~\cite{ref:GM1,ref:GM2} suggests.
A growing mode must decay at large $r$, so its eigenvalue problem does not involve $\mathcal W$ (Appendix~\ref{app:dynamics}), as in the homogeneous analysis of Ref.~\cite{ref:original}. The instability is therefore a property of the solutions under normalizable perturbations, not of the boundary condition imposed on the static source. Since the scale-invariant condition~\eqref{eq:mixedBC} cannot be imposed on such modes (see below), its thermal states with $\alpha>\alpha_*$ are not dynamically stable states, although their equilibrium thermodynamics is stable.

At finite density, on the nonlinear backgrounds, the transverse channel has the shear pole $\Gamma=\eta k^2/(\sE+p)$: the ratio is $0.998$--$0.999$ at $k/T\simeq0.7$--$1.3$ for states with $sT/(\sE+p)$ as small as $0.24$.
In the sound channel the sound and diffusion poles agree with the predictions of linearized hydrodynamics built from $M$, $\zeta$ and $\sigma_Q$ to $0.03$--$1.4\%$ at $k/T\simeq0.7$--$1.6$, with deviations that grow as $k^2$; at $(T/\Lambda,\mu_Q/\Lambda)=(0.134,0.251)$ and $k/T=0.72$, for example, the sound pole is $0.035175-0.002427i$ against $0.035033-0.002431i$, in units $\Lambda=1$.
Just beyond the fold, at $(0.068,0.403)$, where $c_s^2=0.005>0$ and $\det M<0$, the sound pole is damped while the diffusive pole is growing, $\operatorname{Im}\omega=+1.111\times10^{-4}$ against $+1.098\times10^{-4}$ from~\eqref{eq:dir_D}. This confirms that the instability at the fold is diffusive.
At low temperature we could follow only the diffusive pole: it agrees with~\eqref{eq:dir_D} to $0.4\%$ at $\mu_Q/T=28$ and $38$ ($k/T=4$) and to $1.5\%$ at $\mu_Q/T=19$. For the sound pole at $\mu_Q/T\gtrsim10$, and for any pole at $\mu_Q/T\gtrsim100$, our reduction loses accuracy.

\paragraph{Dependence on the ultraviolet condition.}
The decaying condition is the only normalizable one for spacelike momenta, but its relation to a boundary condition of the dual theory has not been derived. We therefore measured how strongly the poles depend on it, by replacing the decaying solution with a mixture containing a multiple $c$ of the growing one.
At $\alpha=0.6$ the relative shift of the shear pole is $0.22\,c\,(k/\Lambda)^6$ and that of the charge-diffusion pole $0.025\,c\,(k/\Lambda)^8$: the hydrodynamic poles at $k\ll\Lambda$ are insensitive to the ultraviolet condition.
In the near zone $\src^{-1/3}\ll r\ll1/(\src k^2)$ the decaying solution contains the source mode of the static dictionary only at relative order $(k/\Lambda)^6$ or higher, so it realizes the Dirichlet condition of the theory with fixed source. This is why its sound pole is that of this theory and not the conformal pole~\eqref{eq:sound_mode}.
The scale-invariant condition~\eqref{eq:mixedBC} would require a source proportional to the condensate, i.e.\ an admixture of the growing solution that is not suppressed at small $k$; we have not found a normalizable formulation of it at $k\neq0$. This is consistent with the view of Ref.~\cite{ref:original} that the condition defines an equilibrium ensemble.
The shear and charge-diffusion poles at zero density involve no scalar source and are the same in both theories.

\paragraph{Zero wavevector.}
At $k=0$ and $\omega\neq0$ the far zone is radiating, and a response function requires a definition of the source. We take it to be the coefficient of the solution that is analytic in $\src$ (the Bessel function $J_4$ in the Maxwell channel), which reduces to the static dictionary as $\omega\to0$. Adding to it a real multiple of the other solution changes only $\operatorname{Im}\sigma$, by a contact term.
With this definition $\operatorname{Re}\sigma(\omega)$ equals the absorption at the horizon. At finite density it reproduces~\eqref{eq:hydro_cond}: $\operatorname{Re}\sigma\to\sigma_Q$ as $\omega\to0$ to four digits ($2.8548$ against $2.8547$ at $\mu_Q/T=1.8$, and $0.41907$ against $0.41902$ at $\mu_Q/T=6.1$), and $\omega\operatorname{Im}\sigma$ gives the Drude weight $\rho^2/(\sE+p)$ to $0.2\%$.
At zero density $\operatorname{Re}\sigma(\omega)$ rises from $\sigma_Q$ to a maximum of $1.5$--$2.2\,\sigma_Q$ at $\omega\simeq\Lambda$ and then falls as $16\pi G_4\operatorname{Re}\sigma\to(2304/\pi)(\Lambda/\omega)^9$ (Figure~\ref{fig:dir_modes}b). This decrease is caused by the growth of the coupling, $Z(\phi_0)\propto r^{9/2}$, in the asymptotic region: the metric there is conformally flat and the Maxwell action is conformally invariant in four dimensions, so a constant coupling would not produce it.
A quantity that does not depend on the definition of the source is the probability that a wave coming in from the far zone is absorbed by the horizon. For $\omega\ll\Lambda$ it is $\mathcal A=(\pi/576)(\omega/\Lambda)^9\,16\pi G_4\operatorname{Re}\sigma$, with which our numerical values agree to $0.5\%$ or better for $\omega\le0.7\Lambda$; it is $10^{-13}$ at $\omega=0.05\Lambda$ and $3\times10^{-3}$ at $\omega=0.7\Lambda$, and it approaches $1$ for $\omega\gtrsim1.7\Lambda$.
The near-horizon region and the radiating region are therefore coupled only at order $(\omega/\Lambda)^9$, which indicates that any additional dissipation from the radiating region is strongly suppressed at low frequency (Section~\ref{sec:transport:sigma}).

\section{Discussion}
\label{sec:discussion}

\subsection{Summary of results}

We have constructed the electrically charged extension of a hairy black brane whose scalar source runs logarithmically, first perturbatively to second order in the charge and then nonlinearly at arbitrary density.
With the scale-invariant boundary condition of Ref.~\cite{ref:original}, imposed at finite charge through the Ward identity~\eqref{eq:full_k}, the second-order thermodynamics is fixed in closed form by the first law and the closed-form susceptibility: the charged Euler relation $3\sE=2TS+2\mu_Q\rho$, the mass correction $\sE^{(2)}|_{S,\alpha}=\mu_Q^{(1)}/(8\pi G_4l)$ and the temperature correction~\eqref{eq:T2_closed}, which the bulk solution reproduces from the surface gravity to $3\times10^{-8}$.
The third-order chemical potential gives the cumulant ratio $T^2\chi_4/\chi_2=6c(\alpha)<0$, and all second-order quantities agree between independent methods to the accuracies of Table~\ref{tab:second_checks}.
The nonlinear branes are locally stable, the pressure is analytic in $\mu_Q^2$ near zero density, no regular extremal horizon exists at finite scalar value, and $\rho/s<\sqrt{21}/(2\pi)$ exactly.
The extremal limit is an infrared throat that is locally $\AdS_2\times\mathbb R^2$, with a scalar running as $\tfrac{\sqrt7}{4}\ln\ln(1/u)$ and an $\AdS_2$ radius growing logarithmically (Proposition~\ref{prop:throat}); the bound is saturated there, the zero-temperature entropy density is finite, and the entropy, $\rho/s$ and the specific heat approach their extremal values as inverse powers of $\ln(\mu_Q/T)$.
Hydrodynamics built on this ensemble, which is established only in equilibrium, has $\eta/s=1/4\pi$, $c_s^2=1/2$ and $\zeta=0$, a charge diffusion constant with $4\pi TD_Q/3=e^{2\alpha}j(\alpha)$ decreasing monotonically from $1$ to $1/4$, and an incoherent conductivity whose charge correction is dominated by a kinematic term.
The boundary potential $\mathcal V=k\sigma^3$ is unbounded below for all hairy branes, so that, if the Hertog--Horowitz criterion applies to the present asymptotics, the thermal states are at best metastable.

In the theory with fixed source the same solutions behave very differently. The susceptibility and the second-order thermodynamics are again in closed form, but the correction to $\chi$ is linear in $T/\Lambda$, and the quartic coefficient $c_{\rm D}$ is positive on the stable branch and diverges at the maximal temperature, a divergence that resums into the shift $T_{\max}(\mu_Q)=T_{\max}[1-0.157\,(\mu_Q/T_{\max})^2]$. The charged branes fill the region below $T_{\max}(\mu_Q)$, which ends at $\mu_{\max}=0.50565\,\Lambda$. Inside it they are locally stable, and on its boundary the coupled heat and charge diffusion mode becomes unstable while sound is still stable. At low density the zero-temperature charge density follows from the infrared fixed point and a single ultraviolet correction to the potential drop, which also explains why $\mu_Q$ is bounded. Buchel's bound is violated at all densities in both of its proposed forms. Quasinormal modes computed with the normalizable condition at large radius reproduce the speed of sound, the bulk viscosity from the Eling--Oz formula, the diffusion constants and the conductivities of this theory, are insensitive to the ultraviolet condition up to corrections of order $(k/\Lambda)^6$, and show that the neutral branes with $\alpha>\tfrac12\ln5$ are dynamically unstable under normalizable perturbations, independently of the boundary condition on the static source, so that the thermal states of the scale-invariant theory with such $\alpha$ are not dynamically stable even though their thermodynamics is.

\subsection{Open problems}

\begin{itemize}
\item \textbf{Holographic dictionary.} Holographic renormalization, the multi-trace interpretation of the boundary condition and the Hertog--Horowitz criterion have been used here as working assumptions (A1)--(A3) (Section~\ref{sec:setup}), supported by internal consistency. A derivation adapted to an asymptotic region that is conformally flat and lies at infinite light-crossing time would put these results on a firmer footing.
\item \textbf{Metastability and dynamics.} The boundary potential is unbounded below (Section~\ref{sec:thermo:stability}), while the homogeneous analysis of Ref.~\cite{ref:original} (see also Appendix~\ref{app:dynamics}) has found no growing modes of the neutral branes. For the theory with fixed source, Section~\ref{sec:dir:qnm} computes the hydrodynamic quasinormal modes; the sound channel near extremality, where the gauge-invariant equations have an apparent singularity, and modes at $\mu_Q/T\gtrsim100$ remain to be computed, and the definition of the source at $k=0$ adopted there has not been derived from first principles. For the scale-invariant condition no dynamical formulation is available. Computing the decay rate of the thermal branes, for instance through the nucleation of regions with large $\sigma$, would quantify their lifetime and clarify whether the theory can be completed, as has been discussed for the analogous deformations of ABJM theory~\cite{ref:CHT}.
\item \textbf{Holographic superconductivity.} The dilatonic Maxwell coupling and the charged background constructed here are natural starting points for holographic superconductor model building~\cite{ref:HHH}.
\end{itemize}

\appendix

\section{Near-horizon structure and the surface gravity}
\label{app:A}

\paragraph{Pole cancellation.}
A horizon shift is generated by the relabeling mode $\partial_m$ of the neutral background at fixed $\alpha$ (together with its compensating diffeomorphism, which is regular at $r_h$).
Its poles in $\delta A$ and $\delta B$ both originate from the term $\partial_m f_0/f_0$, with \emph{opposite} signs, because $\ln a_0^2=\ln f_0+\ln\Omega_0$ contains $+\ln f_0$ while $\ln b_0^2=3\ln\Omega_0-\ln f_0$ contains $-\ln f_0$.
The residues in~\eqref{eq:NHexpand} are therefore exactly opposite: the common pole coefficient of $\partial_m\ln a_0^2$ and $\partial_m\ln b_0^2$ is $\pm\,2\,\delta m/(r_h f_0'(r_h))$, and $\delta r_h=2\,\delta m/(r_h f_0'(r_h))$ by virtue of $f_0'(r_h)=3r_h/l^2$.
Consequently $g=\delta A+\delta B$ is regular, with $g(r_h)=D+D_B$, and $U=f_0\delta A$ is regular, with $U(r_h)=-f_0'(r_h)\delta r_h$ and $U'(r_h)=f_0'(r_h)D$ (using $f_0''(r_h)=0$).

\paragraph{The equation for $g$.}
The $tt$--$rr$ combination of the Einstein equations,
\begin{equation}
R^t{}_t-R^r{}_r=-\tfrac12\,\frac{(\phi')^2}{b^2},
\label{eq:appA_ttrr}
\end{equation}
receives no contribution from a purely electric field, since $T^t{}_t=T^r{}_r$ for the Maxwell stress.
Linearizing in the gauge $\delta H\equiv0$ gives~\eqref{eq:gprime}.
In contrast, the $rr$ equation taken alone contains the electric stress through $g^{rr}\propto 1/b_0^2$, which vanishes at the horizon; written in terms of $U$ and $g$ it yields an equation for $U'$ whose coefficients are finite at $r_h$.

\paragraph{Regularity of the scalar equation.}
The scalar equation reads
\begin{equation}
\frac{1}{r^2\Omega_0^3}\bigl(f_0\,r^2\,\delta\phi'\bigr)' - V''(\phi_0)\,\delta\phi
 = \mathcal S_\phi(r),
\label{eq:appA_scalar}
\end{equation}
where $\mathcal S_\phi$ collects the Maxwell drive $Z'(\phi_0)F^2\propto(A_t^{(1)\prime})^2$ and the couplings to $\delta A$ and $\delta B$.
Expressed through $U$ and $g$, the metric couplings contain at most a simple pole at $r_h$, with a coefficient linear in $(U,g)$. Multiplying by $r_h$-regular factors, the equation takes the form $\delta\phi''=\mathcal N/f_0$ quoted in Section~\ref{sec:regularity:vars}, with $\mathcal N$ regular.
The condition $\mathcal N(r_h)=0$ fixes $\delta\phi'(r_h)$; all other terms are finite, so no further condition arises.
We have verified numerically that the right-hand sides for $U'$ and $g'$ remain finite as $r\to r_h$ and that the $xx$ Einstein equation is satisfied along the solutions.

\paragraph{Surface gravity.}
For $ds^2=-a^2dt^2+b^2dr^2+h^2dx_i^2$ with $a^2(r_*)=0$, Euclidean regularity gives
\begin{equation}
T=\frac{1}{4\pi}\,\frac{(a^2)'(r_*)}{\sqrt{(a^2b^2)(r_*)}} ,
\label{eq:appA_T2}
\end{equation}
where we used that $a_0^2b_0^2=w^2$ is smooth and nonvanishing, so that $a^2b^2=w^2(1+\eps^2g)$ has no pole.
With $a^2=\Omega_0\,(f_0+\eps^2U)$, the horizon sits at $r_*=r_h+\eps^2\delta r_h$ with $\delta r_h=-U(r_h)/f_0'(r_h)$, and
\begin{equation}
(a^2)'(r_*)=\Omega_0(r_*)\bigl[f_0'(r_*)+\eps^2U'(r_h)\bigr],\qquad
(a^2b^2)(r_*)=w(r_h)^2\Bigl[1+\eps^2\Bigl(2\,\delta r_h\,\tfrac{w'}{w}\big|_{r_h}+g(r_h)\Bigr)\Bigr].
\end{equation}
Using $f_0''(r_h)=0$ and $\Omega_0'/\Omega_0|_{r_h}=w'/(2w)|_{r_h}=3e^{-\alpha}\sinh\alpha/r_h$, one obtains~\eqref{eq:T2formula}.
The time coordinate in~\eqref{eq:appA_T2} is the one in which the metric is written. If $\delta A\to\delta A_\infty\neq0$, the boundary time is $\tilde t=(1+\tfrac12\eps^2\delta A_\infty)\,t$ and the temperature must be divided by the same factor. Equation~\eqref{eq:T2formula} is quoted in the normalization $\delta A_\infty=0$.

\section{The extremal infrared throat}
\label{app:throat}

\paragraph{Field equations.}
We use the gauge $ds^2=-D(u)\,dt^2+du^2/D(u)+H(u)^2(dx^2+dy^2)$, in which the Gauss law gives $A_t'=Q/(H^2Z)$, and set $l=1$. With $\Upsilon\equiv e^{-3\phi/\sqrt7}$, so that $e^{-\sqrt7\phi}=\Upsilon^{7/3}$, the combination of the $tt$ and $uu$ Einstein equations, the $uu$ constraint, the $xx$ equation and the scalar equation read
\begin{align}
\frac{H''}{H}&=-\frac{\phi'^2}{4},
\label{eq:app_ttuu}\\
\frac{D'H'}{H}+\frac{DH'^2}{H^2}-\frac{D\phi'^2}{4}&=\frac{21}{4}\Upsilon-\frac94\Upsilon^{7/3}-\frac{Q^2\Upsilon}{H^4},
\label{eq:app_uu}\\
\frac{D''}{2}+\frac{D'H'}{H}&=\frac{21}{4}\Upsilon-\frac94\Upsilon^{7/3}+\frac{Q^2\Upsilon}{H^4},
\label{eq:app_xx}\\
\frac{(H^2D\phi')'}{H^2}&=\frac{9\sqrt7}{2}\bigl(\Upsilon-\Upsilon^{7/3}\bigr)-\frac{6\sqrt7}{7}\frac{Q^2\Upsilon}{H^4}
=\partial_\phi\Bigl(V+\frac{2Q^2}{ZH^4}\Bigr)\Big|_H .
\label{eq:app_s}
\end{align}
We derived them from the action~\eqref{eq:action} symbolically.
They are equivalent to~\eqref{eq:full_F}--\eqref{eq:full_phi} under $r=H$.
With $u=e^{-\tau}$ and $D=e^{-2\tau}P(\tau)$, and dots denoting $d/d\tau$, they become
\begin{align}
&\ddot H+\dot H+\tfrac14H\dot\phi^2=0,\qquad
-(2P-\dot P)\frac{\dot H}{H}+P\frac{\dot H^2}{H^2}-\frac{P\dot\phi^2}{4}=\frac{21}{4}\Upsilon-\frac94\Upsilon^{7/3}-\frac{Q^2\Upsilon}{H^4},\nonumber\\
&\tfrac12\bigl(2P-3\dot P+\ddot P\bigr)-(2P-\dot P)\frac{\dot H}{H}=\frac{21}{4}\Upsilon-\frac94\Upsilon^{7/3}+\frac{Q^2\Upsilon}{H^4},\nonumber\\
&P\ddot\phi+(\dot P-P)\dot\phi+2P\frac{\dot H}{H}\dot\phi=\frac{9\sqrt7}{2}\bigl(\Upsilon-\Upsilon^{7/3}\bigr)-\frac{6\sqrt7}{7}\frac{Q^2\Upsilon}{H^4}.
\label{eq:app_tau}
\end{align}
For $Q^2=\tfrac{21}{4}H^4$ and without the terms in $\Upsilon^{7/3}$, these equations are solved by $H=h_0$, $\phi=$ const and $P=\tfrac{21}{2}\Upsilon$ for every value of $\phi$. This is the flat direction of $\AdS_2\times\mathbb R^2$ solutions discussed in Section~\ref{sec:finite:thermo}.

\paragraph{Leading order.}
Set $H=h_0(1+\eta)$ and assume that $\ln \Upsilon$, $\ln(P/\Upsilon)$ and $\eta$ vary on scales $\Delta\tau\sim\tau$, so that each $\tau$ derivative costs a factor $1/\tau$.
Consider first a general $\varepsilon$, defined in~\eqref{eq:Veff_IR}.
At leading order the $xx$ equation gives $P\simeq(\tfrac{21}{2}-\varepsilon)\,\Upsilon$, and the scalar equation gives $-P\dot\phi\simeq2\delta\varepsilon\,\Upsilon$ with $\delta=3/\sqrt7$, i.e.
\begin{equation}
\dot\phi\simeq-\frac{4\delta\varepsilon}{21-2\varepsilon}.
\label{eq:app_eps}
\end{equation}
For $\varepsilon>0$ the scalar decreases toward the horizon, contradicting $\phi\to\infty$.
For $\varepsilon<0$ it grows linearly in $\tau$.
The first equation of~\eqref{eq:app_tau} with constant $\dot\phi\neq0$ then gives $H\propto e^{\lambda\tau}$ with $\lambda^2+\lambda+\dot\phi^2/4=0$, whose roots have negative real parts, so that $H\to0$.
This contradicts the lower bound on the entropy.
Hence $\varepsilon=0$. The equations then reduce at leading order to
\begin{equation}
P\simeq\frac{21}{2}\,\Upsilon,\qquad
\dot\eta\simeq-\frac{\dot\phi^2}{4},\qquad
21\,\eta\simeq\frac94\,\Upsilon^{4/3},\qquad
-P\dot\phi\simeq18\sqrt7\,\eta\,\Upsilon-\frac{9\sqrt7}{2}\,\Upsilon^{7/3}.
\label{eq:app_LO}
\end{equation}
The third relation comes from the $uu$ constraint, whose left-hand side is of order $\Upsilon/\tau^2$ while $\eta \Upsilon$ and $\Upsilon^{7/3}$ are of order $\Upsilon/\tau$.
The last two give $\dot\phi\simeq\tfrac{12\sqrt7}{49}\,\Upsilon^{4/3}$, and with $\dot \Upsilon=-\delta \Upsilon\dot\phi$ this integrates to $\Upsilon^{4/3}\simeq49/(48\tau)$ up to a shift of $\tau$.
Then $\eta\simeq\tfrac{3}{28}\Upsilon^{4/3}=7/(64\tau)$, and the second relation is satisfied identically: $\dot\eta=-7/(64\tau^2)=-\dot\phi^2/4$ with $\dot\phi=\sqrt7/(4\tau)$.
This establishes~\eqref{eq:throat} at leading order.

\paragraph{Next-to-leading order.}
Substituting
\begin{align}
\Upsilon&=\Bigl(\frac{49}{48\tau}\Bigr)^{3/4}\Bigl(1+\frac{a_1\ln\tau+b}{\tau}\Bigr),\qquad
H=h_0\Bigl(1+\frac{7}{64\tau}+\frac{c_1\ln\tau+d_1}{\tau^2}\Bigr),\nonumber\\
P&=\frac{21}{2}\,\Upsilon\Bigl(1+\frac{e_1\ln\tau+f_1}{\tau}+\frac{g_2\ln^2\tau+e_2\ln\tau+f_2}{\tau^2}\Bigr)
\end{align}
into~\eqref{eq:app_tau} and requiring all equations to hold at the next order in $1/\tau$, including the $\ln\tau$ terms, gives a unique solution for every $b$:
\begin{gather}
a_1=\tfrac{27}{64},\qquad c_1=\tfrac{63}{1024},\qquad d_1=\tfrac{7}{48}\,b+\tfrac{2079}{24576},\nonumber\\
e_1=g_2=0,\qquad f_1=-\tfrac{25}{16},\qquad e_2=-\tfrac{225}{256},\qquad f_2=-\tfrac{5}{12}\,(5b-9).
\label{eq:app_NLO}
\end{gather}
The constant $b$ corresponds to the shift of $\tau$. Without the $\ln\tau$ terms the equations at this order are inconsistent, so the logarithms are required.

\paragraph{Near-extremal thermodynamics.}
At small $T$ the horizon lies at $u_h=e^{-\tau_h}\ll1$. Over $\Delta\tau=\order{1}$ around $\tau_h$, $\Upsilon$ and $H$ change by relative amounts of order $1/\tau_h$ and $1/\tau_h^2$, so the geometry there is the $\AdS_2$ black hole $D\simeq\tfrac{21}{2}\,\Upsilon_h\,(u^2-u_h^2)$.
Its temperature is $D'(u_h)/(4\pi)=21\,\Upsilon_h\,u_h/(4\pi)$ in the throat time, which differs from the boundary time by a $T$-independent factor at leading order.
This gives~\eqref{eq:tau_h}.
The entropy is $s=H(u_h)^2/(4G_4)=s_0\bigl(1+2\eta(\tau_h)\bigr)$ up to corrections of relative order $\ln\tau_h/\tau_h^2$, which gives~\eqref{eq:lowT}. For $C_\rho$ we used $d\tau_h/d\ln(\mu_Q/T)=1+\order{1/\tau_h}$.

\section{Time-dependent perturbations and the asymptotic region}
\label{app:dynamics}

Positive specific heat is necessary but not sufficient for dynamical stability (Section~\ref{sec:thermo:stability}); see Ref.~\cite{ref:Berti-QNM} for a review of quasinormal-mode methods.
The homogeneous sector of the neutral branes, i.e.\ zero-momentum perturbations $\propto e^{-i\omega t}$ of the coupled scalar--gravity channel, was analyzed in Ref.~\cite{ref:original}, which gives the master equation and the numerical argument-principle search and finds no growing modes in the region examined.
Here we only recall the asymptotic structure that matters for the present paper.
Since $b^2/a^2=\Omega_0^2/f_0^2\simeq\src l^4/r$, the frequency term dominates the static terms, which fall off as $r^{-2}$, for $r\gg1/(\src\omega^2)$, where $\delta\phi\simeq r^{-7/4}e^{\pm2i\omega\sqrt{\src}\,l^2\sqrt r}$. This reflects the infinite light-crossing time of the asymptotic region (Section~\ref{sec:setup}).
Time-dependent perturbations can therefore carry energy into the asymptotic region, and the source/condensate split that defines~\eqref{eq:mixedBC} applies only to static data.
A growing mode, $\operatorname{Im}\omega>0$, must be ingoing at the horizon and decaying at large $r$, a condition independent of $\mathcal W$, so the stability question is not affected.
The transport coefficients of Section~\ref{sec:transport} are obtained from horizon data at $\omega\to0$ and do not rely on a prescription for finite-frequency response.
For nonzero wavevector the large-$r$ equations become Bessel equations in $2\sqrt{k^2-\omega^2}\,\src^{1/2}l^2r^{1/2}$, so that for $|\omega|<k$ the waves are evanescent and the radiating region is absent; Section~\ref{sec:dir:qnm} uses this to compute the quasinormal modes of the theory with fixed source and gives a definition of the response at $k=0$.

\section{Numerical convergence}
\label{app:C}

Table~\ref{tab:conv} shows, at the representative value $\alpha=0.6$, the dependence of the numerically constructed solution with unshifted source, i.e.\ of its temperature correction and horizon scalar at fixed $(S,\src)$, on the starting point $x_0$ of the integration, the outer radius $r_{\max}$ and the integrator tolerance.
The relabeling to fixed $\alpha$ uses only the neutral thermodynamics and closed-form integrals, and adds a negligible error.
The corresponding first-law value of the Dirichlet ensemble is $-0.0473416629$.
The residual difference is controlled by the $\order{r^{-3}}$ corrections to the asymptotic constants and decreases accordingly with $r_{\max}$.
Solutions in four different labelings, with horizon data $(U,g,\delta\phi)(r_h)=(0,0,0)$, $(1,0,0)$, $(0,0,1)$ and $(0.3,0.7,-0.2)$, give the same $T^{(2)}/T^{(0)}|_{S,\src}$ at $\alpha=0.6$ after conversion to fixed $(S,\src)$ through the neutral thermodynamics, from $-0.047341662$ to $-0.047341671$.

\begin{table}[ht]
\centering
\caption{Convergence at $\alpha=0.6$ in units $16\pi G_4=l=m=1$, as in Table~\ref{tab:thermo}. Both entries are dimensionless and independent of $G_4$. Unless varied, $x_0=10^{-7}$, $r_{\max}=400$, relative tolerance $10^{-12}$.}
\label{tab:conv}
\renewcommand{\arraystretch}{1.15}
\begin{tabular}{lcc}
\toprule
Parameter & $T^{(2)}/T^{(0)}|_{S,\src}$ & $\delta\phi(r_h)|_{S,\src}$ \\
\midrule
$x_0=10^{-5},\,10^{-6},\,10^{-7}$ & $-0.0473416623$ (all) & $0.004560336$ (all) \\
$r_{\max}=50$ & $-0.0473413725$ & $0.004560343$ \\
$r_{\max}=100$ & $-0.0473416266$ & $0.004560337$ \\
$r_{\max}=400$ & $-0.0473416623$ & $0.004560336$ \\
$r_{\max}=1000$ & $-0.0473416628$ & $0.004560336$ \\
tolerance $10^{-9}$--$10^{-12}$ & $-0.0473416623$ (all) & $0.004560336$ (all) \\
\bottomrule
\end{tabular}
\end{table}

\section*{Acknowledgments}

The author thanks J.~Kim for helpful discussions.

\end{document}